\documentclass[3p]{elsarticle}

\usepackage{nicefrac}
\usepackage{subcaption}
\usepackage{here}
\usepackage{amssymb}
\usepackage{amsmath}
\usepackage{upgreek}
\usepackage{algorithm}
\usepackage{array}

\usepackage{hyperref}

\usepackage{color}
\usepackage[normalem]{ulem}
\newcommand{\Delete} [1]{\bgroup\noindent\textcolor{red}{\xout{#1}}\egroup\ignorespacesafterend}
\newcommand{\Insert} [1]{\bgroup\noindent\textcolor{blue}{#1}\egroup\ignorespacesafterend}

\newcommand{\frage}[1]{{#1}}
\newcommand{\budgetremark}[1]{\frage{\color{brown} [budget: #1]}}
\renewcommand{\budgetremark}[1]{}

\journal{arXiv.org}

\begin{document}

\begin{frontmatter}

\title{Microstructure evolution of compressed micropillars investigated by in situ HR-EBSD analysis and dislocation density simulations}

\author[1add]{Kolja Zoller}

\author[2add]{Szilvia Kal\'{a}cska}

\author[3add]{P\'{e}ter Dus\'{a}n Isp\'{a}novity}

\author[1add,4add]{Katrin Schulz\corref{mycorrespondingauthor}}
\cortext[mycorrespondingauthor]{Corresponding author}
\ead{katrin.schulz@kit.edu}

\address[1add]{Karlsruhe Institute of Technology (KIT), Institute for Applied Materials - Computational Materials Science (IAM-CMS), Kaiserstr. 12, 76131 Karlsruhe, Germany}
\address[2add]{Empa, Swiss Federal Laboratories for Materials Science and Technology, Laboratory of Mechanics of Materials and Nanostructures, CH-3602 Thun, Feuerwerkerstrasse 39. Switzerland}
\address[3add]{E{\"o}tv{\"o}s Lor{\'a}nd University, Department of Materials Physics, P\'azm\'any P. stny. 1/A, 1117 Budapest, Hungary}
\address[4add]{Karlsruhe University of Applied Sciences, Moltkestrasse 30, 76133, Karlsruhe, Germany}

\begin{abstract}

With decreasing system sizes, the mechanical properties and dominant deformation mechanisms of metals change. 
For larger scales, bulk behavior is observed that is characterized by a preservation and significant increase of dislocation content during deformation whereas at the submicron scale very localized dislocation activity as well as dislocation starvation is observed. 
In the transition regime it is not clear how the dislocation content is built up.
This dislocation storage regime and its underlying physical mechanisms are still an open field of research.
In this paper, the microstructure evolution of single crystalline copper micropillars with a $\langle1\,1\,0\rangle$ crystal orientation and varying sizes between $1$ to $10\,\mu\mathrm{m}$ is analysed under compression loading. 
Experimental \emph{in situ} HR-EBSD measurements as well as 3d continuum dislocation dynamics simulations are presented. 
The experimental results provide insights into the material deformation and evolution of dislocation structures during continuous loading. This is complemented by the simulation of the dislocation density evolution considering dislocation dynamics, interactions, and reactions of the individual slip systems providing direct access to these quantities.
Results are presented that show, how the plastic deformation of the material takes place and how the different slip systems are involved. A central finding is, that an increasing amount of GND density is stored in the system during loading that is located dominantly on the slip systems that are not mainly responsible for the production of plastic slip. This might be a characteristic feature of the considered size regime that has direct impact on further dislocation network formation and the corresponding contribution to plastic hardening.

\end{abstract}

\begin{keyword}
Micropillar Compression \sep Size Effect \sep Dislocation based Plasticity \sep HR-EBSD
\end{keyword}

\end{frontmatter}

\renewcommand\div{\operatorname*{div}}
\newcommand\sym{\operatorname*{sym}}
\newcommand\sgn{\operatorname*{sgn}}
\newcommand\tr{\operatorname*{trace}}
\newcommand\R{\mathbb{R}}
\newcommand\pl{\mathrm{pl}}
\newcommand\el{\mathrm{el}}
\renewcommand\d{\,\mathrm{d}}
\newcommand\dir{\mathrm{D}}
\newcommand\neu{\mathrm{N}}
\newcommand\nsp{S}
\newcommand\Gray[1]{\textcolor{gray}{#1}}
\newcommand\Red[1]{\textcolor{red}{#1}}
\def\mm{\,\upmu\mathrm{m}}
\def\ns{\,\mathrm{ns}}
\def\vec#1{\ensuremath{\mathchoice
                     {\mbox{\boldmath$\displaystyle{#1}$}}
                     {\mbox{\boldmath$\textstyle{#1}$}}
                     {\mbox{\boldmath$\scriptstyle{#1}$}}
                     {\mbox{\boldmath$\scriptscriptstyle{#1}$}}}}
\renewcommand{\rho}{\varrho}
\newcommand{\Vector}[2]{\bigg(\begin{matrix}#1\\#2\end{matrix}\bigg)}

\def\bsigma{\text{\boldmath$\sigma$\unboldmath}}
\def\bkappa{\text{\boldmath$\kappa$\unboldmath}}
\def\btkappa{\text{\boldmath$\tilde{\kappa}$\unboldmath}}
\def\tkappa{\tilde{\kappa}}
\def\bchi{\text{\boldmath$\chi$\unboldmath}}
\def\bepsilon{\text{\boldmath$\varepsilon$\unboldmath}}
\def\bgamma{\text{\boldmath$\gamma$\unboldmath}}
\def\bphi{\text{\boldmath$\varphi$\unboldmath}}
\def\brho{\text{\boldmath$\rho$\unboldmath}}
\def\bbeta{\text{\boldmath$\beta$\unboldmath}}
\def\b#1{\mathbf{#1}}

\def\gliss{\mathrm{gliss}}
\def\cross{\mathrm{cross}}
\def\screw{\mathrm{screw}}
\def\edge{\mathrm{edge}}
\def\old{\mathrm{old}}
\def\new{\mathrm{new}}
\def\mult{\mathrm{mult}}

\section{Introduction}
\label{sec:Introduction}

Plastic deformation of bulk metals is usually a smooth process due to the large dislocation content of the crystal. As the specimen size decreases to or below the micrometer scale the properties of deformation change profoundly in many aspects. First of all, as it was shown by compression experiments of single crystalline micron and sub-micron sized pillars (called micro- or nanopillars, respectively), deformation gets localized in distinct slip bands giving rise to an inhomogeneous slip surface \cite{dimiduk2005size}. Secondly, these distinct slip bands appear in an intermittent fashion in the form of sudden strain bursts that lead to jerky and stochastic stress-strain curves \cite{dimiduk2006scale}. Finally, the strength exhibits inverse dependence on the specimen size in this regime, a phenomenon called size effect \cite{volkert2006size}.

The reason for these specific properties of submicron-scale deformation is that the number of dislocations in the volume is rather limited. In bulk samples, upon loading, dislocations move, interact and multiply in the crystal and build up complex structures called dislocation patterns which have a fundamental influence on the mechanical response. In the case of nanopillar compression experiments, however, most of the dislocations can escape the sample before being able to multiply or interact with other dislocations. This process has been termed dislocation starvation \cite{greer2005size, shan2008mechanical} and results in a limited amount of possible dislocation sources where accumulation of plastic strain is possible. In addition, due to the small specimen size, these sources are typically harder to activate compared to their bulk counterparts, giving rise to the so-called source truncation hardening \cite{parthasarathy2007contribution}. Due to the depletion of the dislocation content from the sample strain hardening is often not observed at this scale. These processes have been successfully modelled with discrete dislocation dynamics (DDD) simulations that track the motion of individual dislocation lines \cite{tang2008dislocation, rao2008athermal} and the size effects were explained in terms of weakest link arguments \cite{el2009role, ispanovity2013average, derlet2015probabilistic}.

In this paper we investigate the intermediate, so far less understood, size regime between the bulk and the dislocation starved limits. Here a significant dislocation content is preserved or built up in the sample during deformation but several properties of nanoscale plasticity summarized above still prevail. This \emph{dislocation storage regime} can be observed for micropillars with diameters larger than $\sim$1 $\mu$m. Here we focus on the deformation of face-centered-cubic (fcc) copper single crystals oriented for multiple slip since in such a case dislocation reactions have the strongest influence and are expected to play an important role in dislocation storage. In bulk it is known for decades that under such conditions a cellular pattern develops \cite{staker1972dislocation, prinz1980dislocation, mughrabi1986long} that exhibits fractal character \cite{hahner1998fractal, zaiser1999fractal}. Several experimental methods have been applied to assess how this morphology changes at the micron scale and how that influences the plastic response. From post mortem transmission electron microscopy (TEM) measurements conducted on compressed micropillars Zhao \emph{et al.}~concluded that the transition between bulk and dislocation starved regimes takes place between 1-5 $\mu$m, where the gradual development of a complex cellular structure is seen as the sample diameter increases \cite{zhao2019critical}. According to a box counting analysis the pattern is self-affine with a fractal dimension comparable to that of bulk samples \cite{hahner1998fractal}. With increasing diameter they also observed the decrease of strength and strain burst sizes and the increase of strain hardening \cite{zhao2019critical}. These observations are in line with previous investigations performed on Ni micropillars \cite{norfleet2008dislocation}. To determine the type of the dislocations Kiener \emph{et al.}~applied electron backscatter diffraction (EBSD) and conducted DDD simulations to conclude that a significant amount of geometrically necessary dislocation (GND) density builds up during deformation \cite{kiener2011work}. In situ microdiffraction experiments of Maa\ss{} \emph{et al.}~also reports about the development of an inhomogeneous stress state and a large GND density in Cu pillars under multiple slip \cite{maass2008crystal}. In a similar experiment performed on tensile specimens oriented for single slip Kirchlechner \emph{et al.}~concluded that although slip occurred on the primary slip system, GNDs were generated and stored in a large number on the inactive slip systems \cite{kirchlechner2011dislocation}.

To date, no experimental method is available that could fully resolve the internal microstructure in a micropillar in the dislocation storage regime. During the preparation of TEM lamellae the micropillar is destroyed and some dislocations may leave the thin TEM foil before imaging. X-ray measurements, on the other hand, can be performed in situ and the beam may penetrate through the whole sample, but one can measure only average microstructural properties along the beam. Recently a new methodology of high (angular) resolution EBSD (HR-EBSD) has been established that is capable of determining the GND content to high precision on a surface of a single crystal using a cross-correlation based technique on the recorded Kikuchi patterns \cite{arsenlis1999crystallographic, wilkinson2010determination}. In addition, individual components of the Nye's dislocation density tensor, that characterizes GND density along with the average local Burgers vector direction, can also be obtained which helps in identifying activated slip planes in the crystal \cite{kalacska20203d, della2020101}. This method has been recently applied by Kal\'acska \emph{et al.}\ to investigate the dislocation structures developing in Cu micropillars oriented for multiple slip \cite{kalacska.2020}. Serial focused ion beam (FIB) sectioning was applied and the HR-EBSD measurement was performed subsequently on each surface in order to obtain information in three dimensions (3d) about the microstructure. A well-developed dislocation cell structure was found in deformed pillars but with significantly lower GND density than that of bulk samples which was found to explain the simultaneous observation of strain hardening and size effects. In summary, the experimental investigations suggest the development of a complex dislocation structure in micropillars in the storage regime with a considerable GND content.

The objective of the present paper is to provide further insight on the dislocation microstructure being developed in the dislocation storage regime. We intend to answer the question whether a significant amount of GNDs form upon deformation (although this is not really expected for uniaxial compression) and if so, in which slip systems these dislocations accumulate. To this end, we perform experiments on Cu single crystals oriented for multiple slip to study the formation of the dislocation pattern in situ at different micropillar sizes. We show that, as suggested earlier, with increasing diameter a gradually evolving cellular structure can be observed with a significant locally varying GND content. The Burgers vector analysis suggests that GNDs primarily form on inactive slip systems characterized by low values of the Schmid factor.

In order to explain this somewhat counter-intuitive experimental result we will also perform computational modelling of the microstructure and its influence on the corresponding plastic response. Several approaches are available on various scales to model the evolution of the internal dislocation structure. Molecular dynamics and DDD simulations have been used extensively to understand the specific features of nanoscale plasticity summarized above \cite{weinberger2008surface, cao2008sample, xu2013molecular, csikor2007dislocation, senger.2008, tang2008dislocation, rao2008athermal, ryu2013plasticity, el2015unravelling, stricker.2018}. However, limitations in computational power prohibits their use at pillar sizes and strain rates investigated in this paper. One, thus, cannot model the evolution of the system on atomic scales or at the level of individual dislocations but needs to use tools on higher scales. In this paper we employ continuum dislocation dynamics (CDD) simulations that consider the evolution of the system in terms of smooth density fields of discrete dislocations. One of the first CDD models was developed by Groma and co-workers in two dimensions (2d) and was derived using statistical physics concepts from the mobility law of individual straight dislocations \cite{groma.2003, groma2015scale}. Although these 2d models are able to capture dislocation patterning of straight edge dislocations in single slip \cite{groma2016dislocation, ispanovity2020emergence}, by construction they cannot account for, e.g., dislocation reactions, cross slip and multiplication, mechanisms that are expected to play a key role under multiple slip conditions. In 3d, inspired by the pioneering work of Groma, Hochrainer and co-workers formulated a CDD theory that is based on the kinematics of curved dislocation lines and, as such, takes into account dislocation curvature and the spatial evolution of homogenized dislocation microstructures  \cite{hochrainer.2014, hochrainer2015multipole, hochrainer2016thermodynamically}. To properly describe the dynamics of the system and the interaction between multiple slip systems, a flux-based discontinuous Galerkin scheme has been introduced in \cite{schulz.2019} and stress terms related to specific dislocation processes have to be introduced in a phenomenological manner. Recently, such formulations were developed to account for cross-slip and glissile dislocation reactions \cite{sudmanns.2019} and dislocation sources \cite{schmitt2019mechanism}. The comparison of the results with DDD simulations showed that the proposed scheme yields a physically correct representation of the interplay between dislocation densities on different slip systems. As such, the 3d CDD model has become capable of being successfully applied to realistic deformation problems, like torsion of microwires \cite{zoller.2020}.

The paper is organised as follows. In section 2 the experimental methods related to micropillar fabrication, subsequent compression, and determination of the internal GND content as well as the basics of the CDD simulations are described in detail. Results obtained by experiments and simulations on the stress-strain response, size dependence and the internal microstructure are presented in Section 3. This is followed in section 4 by an in-depth discussion of the dislocation mechanisms found to play a key role at the investigated size regime and their influence on the observed deformation behaviour. The paper concludes with a summary on remaining open questions and an outlook to future research directions.


\section{Methods}
\label{sec:Methods}

\subsection{Considered system}
\label{ssec:2a_System}
\begin{figure*}[!ht]
    \centering
    \includegraphics[width=0.198\textwidth]{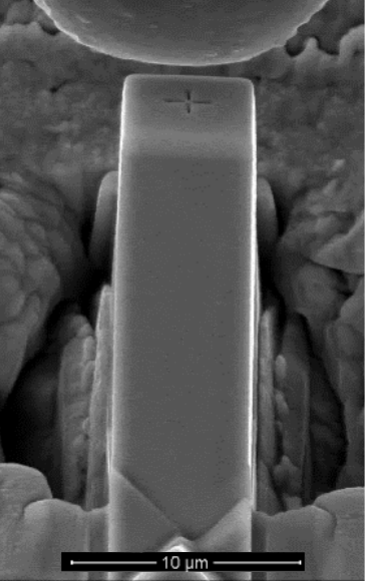}
    \hspace{0.5cm}
    \includegraphics[width=0.7\textwidth]{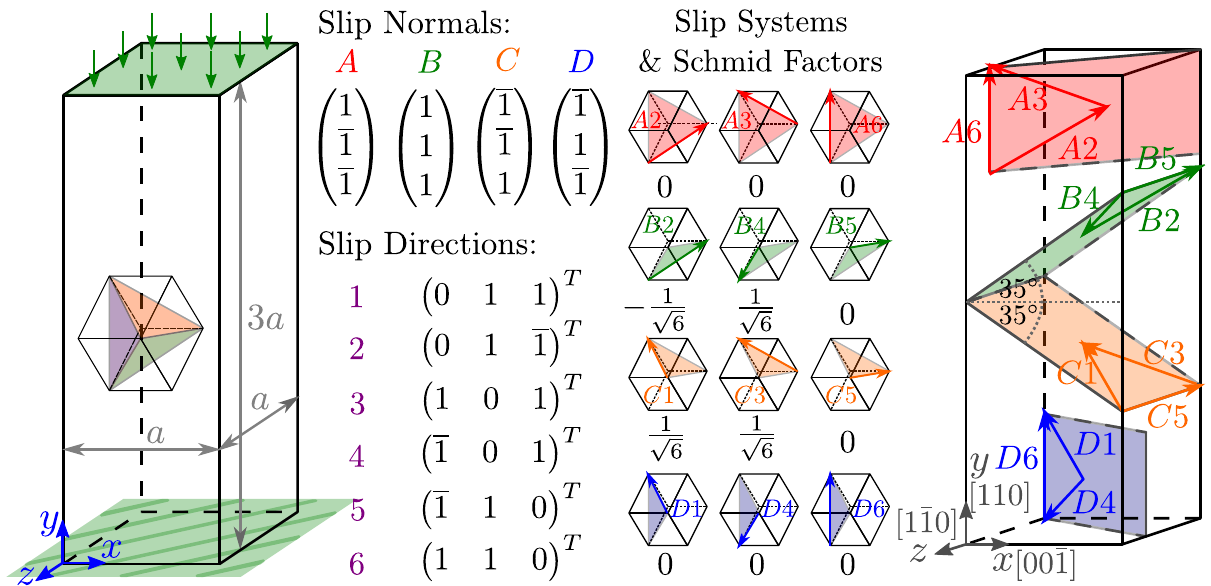}
    \caption{Representation of the real and modelled micropillars as well as the description of the slip systems and their Schmid factors for an ideal uniaxial stress state ($\sigma_{yy}$). Throughout the paper the results will be presented using the $xyz$ coordinate system attached to the micropillar and shown in the sketch. Note that this is different from the crystal coordinate system which is rotated as shown by the given Thompson tetrahedron. 
    }
    \label{fig:2a_system}
\end{figure*}

In order to get insights into the deformation mechanisms and the underlying microstructure evolution of fcc single-crystals at the micron scale, this paper focuses on the compression of copper micropillars oriented for multiple slip by a $\langle1\,1\,0\rangle$ crystal orientation as shown in Figure \ref{fig:2a_system}.
The micropillars have varying sizes in the $a=1-10$  $\mu$m regime
and an aspect ratio of height:width between 2:1 and 3:1 (the latter being shown in Figure \ref{fig:2a_system}). In addition to the crystal coordinate system (represented by the Thompson tetrahedron in Figure \ref{fig:2a_system}) we introduce an $xyz$ coordinate system attached to the micropillar as shown in Figure \ref{fig:2a_system}. Both experimental and simulation results are to be presented in this frame throughout the paper.
The slip systems are named according to the Schmid-Boas notation \cite{schmid.1935} with letters ($A,B,C,D$) for the slip planes and numbers for the slip directions. 

%
Considering the Schmid factors, four slip systems $\{B2,B4,C1,C3\}$ can be identified to play a primary role for a $\langle1\,1\,0\rangle$ crystal orientation, while the other slip systems are expected to be mostly inactive. 
Dislocations on primary slip systems may be able to interact with each other by Lomer and Hirth reactions additionally to coplanar and self-interactions. 
Expecting the same amount of plastic slip on all four primary slip systems would result in a plastic distortion tensor containing only positive $\vec e_{x}\otimes \vec e_{x}$ and negative $\vec e_{y}\otimes \vec e_{y}$ components.
Consequently, the trace of plastic distortion disappears and the volume is preserved.
Considering possible interactions of inactive slip systems, i.e. $\{A2,A3,A6,B5,C5,D1,D4,D6\}$, with the expected primary slip systems $\{B2,B4,C1,C3\}$, the inactive slip systems can be divided into three groups: $\{B5,C5\}$ sharing the same slip planes, $\{A2,A3,D1,D4\}$ sharing the same Burgers vectors, and $\{A6,D6\}$ sharing neither the slip planes nor the Burgers vectors.


\subsection{Experiments}
\label{ssec:2b_Experiments}

For the \emph{in situ} experiments a previously heat-treated copper single crystal sample was used \cite{kalacska.2020}, where an orientation was set in accordance with the previous section. A sharp perpendicular edge of $[1\bar{1}0]$ orientation was prepared by low-energy Ar ion milling at 6.5 kV, 2.4 mA using a Leica EM TIC 3X polisher. 
FIB milling was then applied to create square shape micropillars close to the edge in lathe milling position \cite{uchic.2009} using a Tescan Lyra3 scanning electron microscope (SEM). In order to reduce Ga$^+$ ion implantation to the surface, FIB voltages and currents were subsequently reduced from 30 kV, 10 nA to 5 kV 60 pA as the polishing got closer to the final pillar shape. This way pillars with approximate sides of $1-1.5$ $\mu$m, 3 $\mu$m, 6 $\mu$m and 10 $\mu$m were created with negligible tapering and height:width ratio of 2:1 to reduce bending upon loading. Furthermore, several pillars were prepared with the height:width ratio of approx. 3:1.

Mechanical tests were performed using an Alemnis \emph{in situ} nanoindentation platform (Alemnis AG Switzerland) \cite{wheeler.2013}. The \emph{in situ} frame allowed HR-EBSD mapping while the pillars were kept under load. The targeted maximum deformation for larger pillars ($3-10$ $\mu$m) was $10 \%$, whereas the smallest pillars were deformed up to $\sim 5-10\%$ with strain rates smaller than $10^{-3}\,\nicefrac{1}{\mathrm{s}}$. Conductive diamond flat punch tips ranging from 5 $\mu$m to 25 $\mu$m in diameter were used during the experiments. Mechanical data collected during the compression experiments were analysed by Alemnis AMMDA software. Appropriate corrections were made on the original load-displacement data, including pillar sink-in correction by the Sneddon method \cite{Sneddon.1946}.

EBSD diffraction patterns were recorded with an Oxford Instruments Symmetry detector (AZtec v4.2 software) and an Edax DigiView camera (OIM Data Collection v7 software). During EBSD mapping an electron beam of 20 kV, 10-12 nA was used. HR-EBSD evaluation was carried out with the BLG Vantage CrossCourt v4 software. Cross-correlation image analysis was performed on 20 regions of interest chosen from all diffraction patterns. The  components of the Nye dislocation density tensor (see below) were calculated by a C++ program written by the authors. 

In the paper the Nye dislocation density tensor $\vec \alpha$ \cite{Nye1953} is utilized to characterize the GNDs, defined as


\begin{eqnarray}\label{eq:01}
\alpha_{ij} = \sum_t b_i^t l_j^t\rho^t, \quad (i,j)=x,y,z,
\end{eqnarray}
where dislocations are characterized by their Burgers vector $\vec b^t$ and line direction $\vec l^t$ for different $t$ types of dislocations.
The sum is over all types of dislocations present in the sample, and $\rho ^t$ denotes the density of dislocations of type $t$. 

As a result of the HR-EBSD measurement one obtains the nine independent components of the elastic distortion $\vec \beta^\mathrm{el}$. The values measured at every single point in a $70^{\circ}$ sample tilt typical for HR-EBSD correspond to an average in a surface of around $40\times120$ nm$^2$ and the detected backscattered electrons originate from the first $\sim 50-100$ nm region below the surface \cite{Chen2011}. Due to the chosen sample geometry, HR-EBSD measurements could only be performed on the $[1\bar{1}0]$ surface of the micropillars (that is, the $z=0$ plane), and thus components of $\vec \beta^\mathrm{el}$ were only available as a function of $x$ and $y$ (using the coordinate system of Figure \ref{fig:2a_system}). The Nye tensor $\vec \alpha$ is linked to the elastic distortion $\vec \beta^\mathrm{el}$ as
\begin{eqnarray}\label{eq:aij}
\alpha_{ij} = \left( \mathrm{curl} \vec \beta^\mathrm{el} \right)_{ij} = - \epsilon_{klj} \partial_k \beta^\mathrm{el}_{il},
\end{eqnarray}
where $\mathrm{curl}$ describes the postcurl operator \cite{das.2018}, $\epsilon_{ijk}$ denotes the Levi-Civita symbol and Einstein summation rule is assumed. Since the HR-EBSD measurement only yields data in the $xy$ plane and derivation cannot be performed along the $z$ direction, only three components of the $\vec \alpha$ tensor can be determined experimentally, namely \cite{Wilkinson2009}:
\begin{eqnarray}\label{eq:ai3}
\alpha_{iz} = \partial_y \beta^\mathrm{el}_{ix} - \partial_x \beta^\mathrm{el}_{iy},\quad i=x,y,z.
\end{eqnarray}

Another procedure to get a lower bound estimation for the GND density can be done by utilizing an optimization method to minimize the total dislocation line energy ($L^1$ optimization scheme) \cite{Wilkinson2010}. Terms that cannot be measured are set to zero, and only pure edge or screw dislocations with the same magnitude of Burger’s vector are considered to be present in the sample. These estimations lead to distinguish edge and screw dislocations based on their energies:

\begin{eqnarray}\label{eq:02}
\frac{E_\mathrm{edge}}{E_\mathrm{screw}}=\frac{1}{1-\nu},
\end{eqnarray}
where $\nu$ is the Poisson number. 

Furthermore, we define $\alpha_\mathrm{sq}=\sqrt{\alpha_{xz}^2 + \alpha_{yz}^2 + \alpha_{zz}^2}$ which is proportional to the local GND density \cite{kalacska.2020}. In this article, both $L^1$ and $\alpha_\mathrm{sq}$ methods are used in order to study GND density distribution and to perform Burgers vector analysis in deformed copper single crystalline micropillars.

\subsection{Dislocation density based continuum model}
\label{ssec:2c_Simulations}

The CDD formulation used is based on the framework introduced in \cite{schulz.2019}. Following \cite{sudmanns.2020}, we distinguish between a mobile and a network dislocation density. We take into account multiplication processes including glissile reaction and cross-slip according to \cite{sudmanns.2019} as well as interaction due to Lomer and collinear reactions. 
 This is complemented by the homogenized dislocation source model introduced in \cite{zoller.2020}.
Although several parts of the CDD formulation are known from the literature mentioned above, the used formulation including the considered stress interaction terms, dislocation reactions and the mobility law is presented in this subsection for a better readability.

The modeling approach incorporates two coupled problems: The elastic (external) problem, calculating the elastic stress field for a given deformation state, and the internal problem, describing the microstructure evolution for a given stress field yielding plastic deformation. The two problems are coupled via the plastic slip.

We describe the elasto-plastic deformation by an additive decomposition of the distortion tensor $\mathrm D\b u$ into an elastic part $\bbeta^{\mathrm{el}}$ and a plastic part $\bbeta^{\mathrm{pl}}$. 
Considering small deformations, the strain $\vec\varepsilon$ comprises the symmetric part of the distortion tensor and can be decomposed analogously into an elastic part $\vec \varepsilon^\el$ and a plastic part $\vec \varepsilon^\pl$.
The stress-strain relation describes physical linearity using the Cauchy stress tensor $\vec\sigma$ and the elasticity tensor $\mathbb C$. It holds:
\begin{subequations}
\begin{equation}\tag{\theequation a-c}
\mathrm D\b u = \bbeta^{\mathrm{el}} + \bbeta^{\mathrm{pl}},\hspace{2cm}   \quad   
\vec\varepsilon=\sym D\b u,\hspace{2cm}   \quad   
\vec\sigma =  \mathbb C[\vec \varepsilon - \vec \varepsilon^\pl ].    
\end{equation}
\end{subequations}

The macroscopic equilibrium equation considering body forces $\vec f_{\mathcal B}$ in the continuum $\mathcal B$ is complemented by the boundary conditions at the surface $\partial  \mathcal B$ via given displacements $\vec u_\dir$ for the displacements $\vec u$ on Dirichlet boundaries $\partial_\dir  \mathcal B$ and given traction $\vec t_\neu$ for the applied traction $\vec\sigma \vec n$ on Neumann boundaries $\partial_\neu  \mathcal B$. Dirichlet and Neumann boundaries are disjoint ($\partial_\dir  \mathcal B \cap \partial_\neu  \mathcal B = \varnothing$) and cover the surface completely ($\partial_\dir  \mathcal B \cup \partial_\neu  \mathcal B = \partial  \mathcal B$):
\begin{subequations}
\begin{equation}\tag{\theequation a-c}
-\div\vec\sigma = \vec f_{\mathcal B} \quad \text{ in }\mathcal B
,\hspace{2cm}   \quad   
\vec u = \vec u_\dir  \text{ on } \partial_\dir  \mathcal B
,\hspace{2cm}   \quad
\vec\sigma \vec n = \vec t_\neu \text{ on }   \partial_\neu  \mathcal B.
\end{equation}
\end{subequations}

The plastic distortion is assumed to result solely from
the evolution of the dislocation state on each slip system $s \in \{1,...,S\}$, whereby $S$ describes the total number of considered slip systems (i.e. 12 in the fcc material).
The single slip systems are characterized by their normal vector $\b m_s$, their slip direction $\b d_s=\frac{1}{b_s}\b b_s$ and the line direction of positive edge dislocations $\b l_s = \b m_s \times \b d_s$ forming an orthonormal basis $\{\b d_s,\b l_s,\b m_s\}$.
Herein, $b_s=\Vert\b b_s\Vert$ is the length of the Burgers vector $\b b_s$.
The product of the plastic shear strains $\gamma_s$ on the individual slip systems and the respective Schmid tensor $\b M_s$ is summed up over all slip systems and forms the plastic distortion.
Orowan's equation \cite{orowan.1934} provides the evolution equation for the plastic shear strain based on the dislocation velocity $v_s$ and the mobile dislocation density $\rho^M_s$ on the individual slip systems:
\begin{subequations}
\begin{equation}\tag{\theequation a-c}
\bbeta^{\mathrm{pl}} = \sum_{s=1}^S \gamma_s \b M_s ,\hspace{2cm}   \quad   
\b M_s = \b d_s\otimes\b m_s,\hspace{2cm}   \quad 
\dot\gamma_s = v_s b_s \rho^M_s.    
\end{equation}
\end{subequations}

Here, it is noted that the order in the notation of the plastic distortion tensor is a matter of convention and  both $\b d_s\otimes\b m_s$ or $\b m_s\otimes\b d_s$ are used in the literature.
Since the HR-EBSD method measures the elastic deformation, the elastic part of the distortion tensor is also used for the Nye tensor calculation in the simulation according to equation \ref{eq:aij}. 

\begin{figure*}[!ht]
     \centering
     \includegraphics[width=0.8\linewidth]{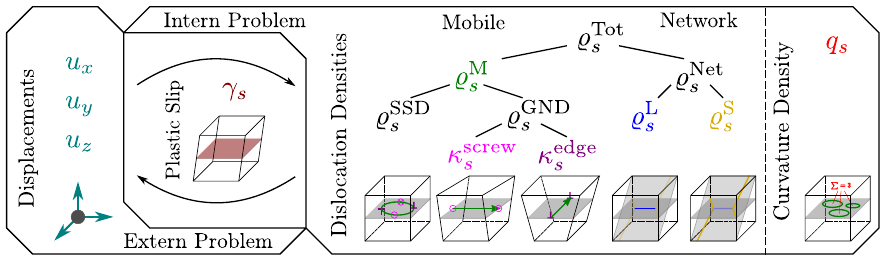}
     \caption{
     CDD formulation with the coupled external and internal problems.
     The internal CDD variables are illustratively shown, whereby the degrees of freedom are highlighted in color. 
     }
     \label{fig:DOF}
\end{figure*}

The dislocation microstructure is described on each slip system by different dislocation densities, specifying the respective dislocation line length per averaging volume, in addition to a curvature density $q_s$. The volume integral of $q_s$ can be interpreted as the number of dislocation loops within the averaging volume. Thereby, the total dislocation density $\rho^\mathrm{Tot}_s$ is subdivided additively into the mobile dislocation density $\rho^\mathrm{M}_s$, which contributes to plastic slip, and the so-called network dislocation density $\rho^\mathrm{Net}_s$, which is sessile. The mobile dislocation density can again be additively decomposed into a statistically stored dislocation (SSD) density $\rho^\mathrm{SSD}_s$ and a vector of the geometrically necessary dislocation (GND) density $\vec \kappa_s$, which forms the GND density $\rho^\mathrm{GND}_s= \Vert \vec \kappa_s \Vert $. The vector of GND density can be written based on its screw and edge part: $\vec \kappa_s = \kappa^\mathrm{screw}_s\b d_s+ \kappa^\mathrm{edge}_s\b l_s$.
The network dislocation density consists of a density of Lomer junctions $\rho^L_s$, which contains the line length of the junction between both end-nodes, and a stabilized dislocation density $\rho^S_s$, which captures the line length of the stable dislocation network due to their attachment to Lomer junctions.
Here, the density of Lomer junctions is shared with the second involved slip system resulting in a prefactor of 0.5 for the calculation of the network dislocation density: $\rho^\mathrm{Net}_s=0.5\,\rho^L_s + \rho^S_s$.
Accordingly the total dislocation density can be decomposed in:
$\rho^\mathrm{Tot}_s=\rho^\mathrm{SSD}_s + \sqrt{(\kappa^\mathrm{screw}_s)^2+(\kappa^\mathrm{edge}_s)^2} + 0.5\,\rho^L_s + \rho^S_s$.
The internal CDD variables are illustratively shown in Figure \ref{fig:DOF}.

The evolution of the CDD quantities defining the averaged dislocation state within the continuum includes the flux-based kinematic description as well as internal dislocation reactions involving different slip systems. This may result in dislocation multiplication as well as increase, transfer or decrease of dislocations line length within the system.
The homogenized, mechanism based dislocation source model according to \cite{zoller.2020} considers the production of new dislocations loops. 
Dislocation sources on the individual slip systems can be activated locally if the effective stress on the individual slip system exceeds the respective critical source stress that depends on the current microstructure.
Multiplication mechanisms are considered by cross-slip and glissile reactions, that can lead to a complex dislocation mobility between different slip systems whereas Lomer reactions form sessile dislocation junctions stabilizing the dislocation network and therefore typically impede the motion of the involved dislocations. The latter process is reversible and the sessile dislocation network can become mobile due to unzipping of the Lomer junctions.
In contrast, the collinear reactions lead to an irreversible decrease of dislocation density due to an annihilation of collinear dislocation lines sections.
The evolution equations of the internal CDD variables read:
\begin{subequations}
\begin{equation}
\partial_t \rho^M_s = - \nabla \cdot \left(v_s\vec\kappa_{s}^\perp \right) + v_s q_s 
+ \partial_t \bar{\rho}^{M,\mathrm{gliss}}_s + \partial_t \bar{\rho}^{M,\mathrm{cross}}_s
+ \partial_t \hat{\rho}^{M}_s + \partial_t \bar{\rho}^{M,\mathrm{Lomer}}_s 
+ \partial_t \rho^{M,\mathrm{react}}_s + \partial_t \rho^{M,\mathrm{cross}}_s
\end{equation}
\begin{equation}
\partial_t \vec\kappa_s = \nabla \times\left(\rho_s v_s \vec m_s\right) + \partial_t \bar{\vec\kappa}^\mathrm{cross}_s
\end{equation}
\begin{equation}
\partial_t \rho^L_s = \partial_t \hat{\rho}^{L}_s - \partial_t \rho^{M,\mathrm{Lomer}}_s - \partial_t \rho^{S,\mathrm{Lomer}}_s
\end{equation}
\begin{equation}
\partial_t \rho^S_s = \partial_t \hat{\rho}^{S}_s - \partial_t \bar{\rho}^{M,\mathrm{Lomer}}_s + \partial_t \rho^{S,\mathrm{react}}_s
\end{equation}
\begin{equation}
\partial_t q_s = -v_s\nabla \cdot \Big(\frac{q_s}{\rho_s}\vec\kappa_s^\perp\Big) - \vec A_s\cdot\nabla^2 v_s 
+ \partial_t \bar{q}^\mathrm{gliss}_s + \partial_t \bar{q}^\mathrm{cross}_s 
+ \partial_t q^\mathrm{react}_s + \partial_t q^\mathrm{cross}_s + \partial_t q^\mathrm{prod}_s.
\end{equation}
\label{density_equation}
\end{subequations}

Here, $()^\mathrm{react}$ is a shortened notation for specific dislocation reactions: $()^\mathrm{react}=()^\mathrm{gliss}+()^\mathrm{Lomer}+()^\mathrm{coll}$. Further, any mechanism that results in the formation of a dislocation reaction is indicated by $\bar{()}$, whereas the unzipping of Lomer junctions is indicated by $\hat{()}$ and $\{\partial_t \rho^{M,\mathrm{react}}_s,\,\partial_t \rho^{M,\mathrm{cross}}_s,\,\partial_t \rho^{S,\mathrm{react}}_s\}$ covers the decrease of dislocation density on the reacting slip systems due to the mechanisms.
For a more detailed consideration of the individual terms, please refer to the literature.

Nevertheless, it should be noted at this point that 
the link length of the stabilized dislocations due to Lomer junctions $L^\mathrm{Lomer}$, defining the strength of these junctions \cite{rodney.1999,shin.2001}, is approximated 
by a Rayleigh distribution 
in a certain interval. 
This interval is based on the scattering of
the nucleation radii and the travel distances of the dislocations measured within dislocation
network of DDD simulations \cite{stricker.2018}. Furthermore, the averaged link length $L^\mathrm{Lomer}_\mathrm{avg}$ is assumed to scale with the averaged dislocation distance 
and is used as the expected value of the Rayleigh distribution.
The critical stress for unzipping of Lomer junctions $\tau^{\mathrm{crit,Lomer}}_s$ is based on the bow out stress of the dislocation links of Lomer junctions and scales inversely to $L^\mathrm{Lomer}$. 
The critical source stress $\tau^{\mathrm{crit}}_s$ responsible for the formation of new dislocations within the system comes mainly into play for cases of low dislocation densities and therefore considers the critical resolved shear stress $\tau^{\mathrm{CRSS}}$, which is taken from the literature.
It holds:
\begin{subequations}
\begin{equation}
\tau^{\mathrm{crit,Lomer}}_s = \frac{\mu b_s}{2} \frac{1}{L^\mathrm{Lomer}}
,\quad
L^\mathrm{Lomer} \in \left\{0.1\,L^\mathrm{Lomer}_\mathrm{avg},\,2\,L^\mathrm{Lomer}_\mathrm{avg}\right\}
,\quad
L^\mathrm{Lomer}_\mathrm{avg} = \frac{1}{\sqrt{\sum_{\tilde{s}=1}^S \rho^\mathrm{Tot}_{\tilde{s}}}}
\end{equation}
\begin{equation}
\tau^{\mathrm{crit}}_s = \max \left\lbrace \frac{\mu b_s}{c^\mathrm{FR}}\sqrt{\sum_{\tilde{s}=1}^S \rho^\mathrm{Tot}_{\tilde{s}}} \,,\, \tau^{\mathrm{CRSS}} \right\rbrace
    ,\quad
   c^\mathrm{FR} > 0 \; .
\end{equation}
\end{subequations}

Using the closure assumptions introduced in \cite{hochrainer.2014}, the dislocation alignment tensor $\vec A_s$ reads
\begin{equation}
    \vec A_s =\frac{1}{2}\bigg(\big(\rho^M_s+\Vert\vec\kappa_{s}\Vert\big)\frac{\vec\kappa_{s}}{\Vert\vec\kappa_{s}\Vert}\otimes \frac{\vec\kappa_{s}}{\Vert\vec\kappa_{s}\Vert}
  + \big(\rho^M_s-\Vert\vec\kappa_{s}\Vert\big)\frac{\vec\kappa_{s}^\perp}{\Vert\vec\kappa_{s}^\perp\Vert}\otimes\frac{\vec\kappa_{s}^\perp}{\Vert\vec\kappa_{s}^\perp\Vert}
  \bigg)
  \qquad \text{,} \qquad  
\bkappa^\bot_s = \bkappa_s\times\b m_s
.
\end{equation}

Regarding the system boundaries, we assume that the dislocations can leave the continuum unhindered at the surfaces, characterized by the surface normal $\vec n$, while no new dislocations are allowed to enter the system over the surface. Accordingly, the boundary conditions are described by Robin-type boundary conditions for the mobile dislocation densities, whereas the curvature density vanishes on the inflow boundaries $\partial_\mathrm{in}  \mathcal B = \left\{ \b x \in \partial \mathcal B: \frac{v_s}{\rho^M_s}\b n \cdot \bkappa^\bot_s < 0\right\}$ and is dissipated on the outflow boundaries $\partial_\mathrm{out} \mathcal B$.
Inflow and outflow boundaries are disjoint ($\partial_\mathrm{in}  \mathcal B \cap \partial_\mathrm{out}  \mathcal B = \varnothing$) and cover the surface completely ($\partial_\mathrm{in}  \mathcal B \cup \partial_\mathrm{out}  \mathcal B = \partial  \mathcal B$). Thus, it holds:
\begin{subequations}
\begin{equation}\tag{\theequation a-b}
v_s\Big(\Vert\vec n\times \vec m_s\Vert\rho^M_s + (\vec n\times \vec m_s)\cdot \vec\kappa_s \Big) = 0 \text{ on }   \partial \mathcal B
,\hspace{2cm}   \quad   
q_s = 0  \text{ on } \partial_\mathrm{in}  \mathcal B
\end{equation}
\end{subequations}

A constitutive law for the velocity is given as closure of the dislocation problem. Using a material specific drag coefficient $B>0$, the effective stress $ \tau_s^\mathrm{eff}$ and the yield stress $ \tau_s^\mathrm{y}$, it reads:
\begin{subequations}
\begin{equation}\tag{\theequation a-b}
  v_s = \frac{b_s}B \sgn\left(\tau_s^\mathrm{eff}\right)\max\Big\{0,|\tau_s^\mathrm{eff}| - \tau_s^\mathrm{y}\Big\}
  ,\quad 
  \tau_s^\mathrm{y} = \mu b_s\sqrt{\sum_{\tilde{s}=1}^S a_{s\tilde{s}} \cdot
  \begin{cases}
     (0.5\,\rho^L_{\tilde{s}}), & \text{for } s \leftrightarrow \tilde{s} = \text{Lomer} \\
     (\rho^M_{\tilde{s}} + \rho^S_{\tilde{s}}), & \text{otherwise}
  \end{cases}
  }.
\end{equation}
\end{subequations}
The yield stress accounts for hardening due to the interaction with forest dislocations from other slip systems within the averaging volume.
Here, we consider the yield stress term based on \cite{franciosi.1980} and adapt it following \cite{sudmanns.2020} including a material interaction matrix $a_{s\tilde{s}}>0$ and the shear modulus $\mu$.
The effect of elastic anisotropy is averaged out in large dislocation densities and consequently the isotropic shear modulus can be used, see \cite{madec.2003}.
The variable $\tilde{s}$ describes the index of summation over the individual slip systems.
%
%
Furthermore, the resolved shear stress on a slip system $\tau_s$ and the back stress $\tau_s^\mathrm{b}$ contribute to the effective stress and are given by:
\begin{subequations}
\begin{equation}\tag{\theequation a-c}
\tau_s^\mathrm{eff} = \tau_s - \tau_s^\mathrm{b}
,\hspace{1.5cm}   \quad 
\tau_s = \vec\sigma \cdot \vec M_s
,\hspace{1.5cm}   \quad   
\tau_s^\mathrm{b} = \frac{D\mu b_s}{\rho^M_{s}} \nabla \cdot \vec\kappa_{s}^\perp \quad\mathrm{,}\quad D=\frac{3.29}{2 \pi^2(1-\nu)}.
\end{equation}
\end{subequations}

The resolved shear stress is determined by the projection of the stress tensor including dislocation eigenstresses on each slip system following \cite{schulz.2014}. The back stress covers short-range dislocation stress interaction according to \cite{groma.2003,schmitt.2015} depending on a material parameter $D$, or respectively the Poisson's ratio $\nu$, and the shear modulus $\mu$.

\subsubsection*{Numerical implementation}
The CDD formulation has been implemented into a two-scale numerical framework based on \cite{schulz.2019} using an own customized version of the parallel finite element software M++ \cite{wieners.2010,wieners.2005}. Thereby, a finite element approach is used with hexahedral elements and linear ansatz functions for the displacements in the external problem as well as constant ansatz functions and an implicit Runge-Kutta discontinous Galerkin scheme with full upwind flux for the densities in the internal problems. Simplifying but found to be efficient, the same mesh resolution is used for both scales. The time is discretized by an implicit midpoint rule with a fixed time step.

\section{Results}
\label{sec:Results}

\subsection{Experiments}
\label{ssec:3b_Experiments}

\begin{figure*}[!ht]
    \centering
    \includegraphics[width=0.70\textwidth]{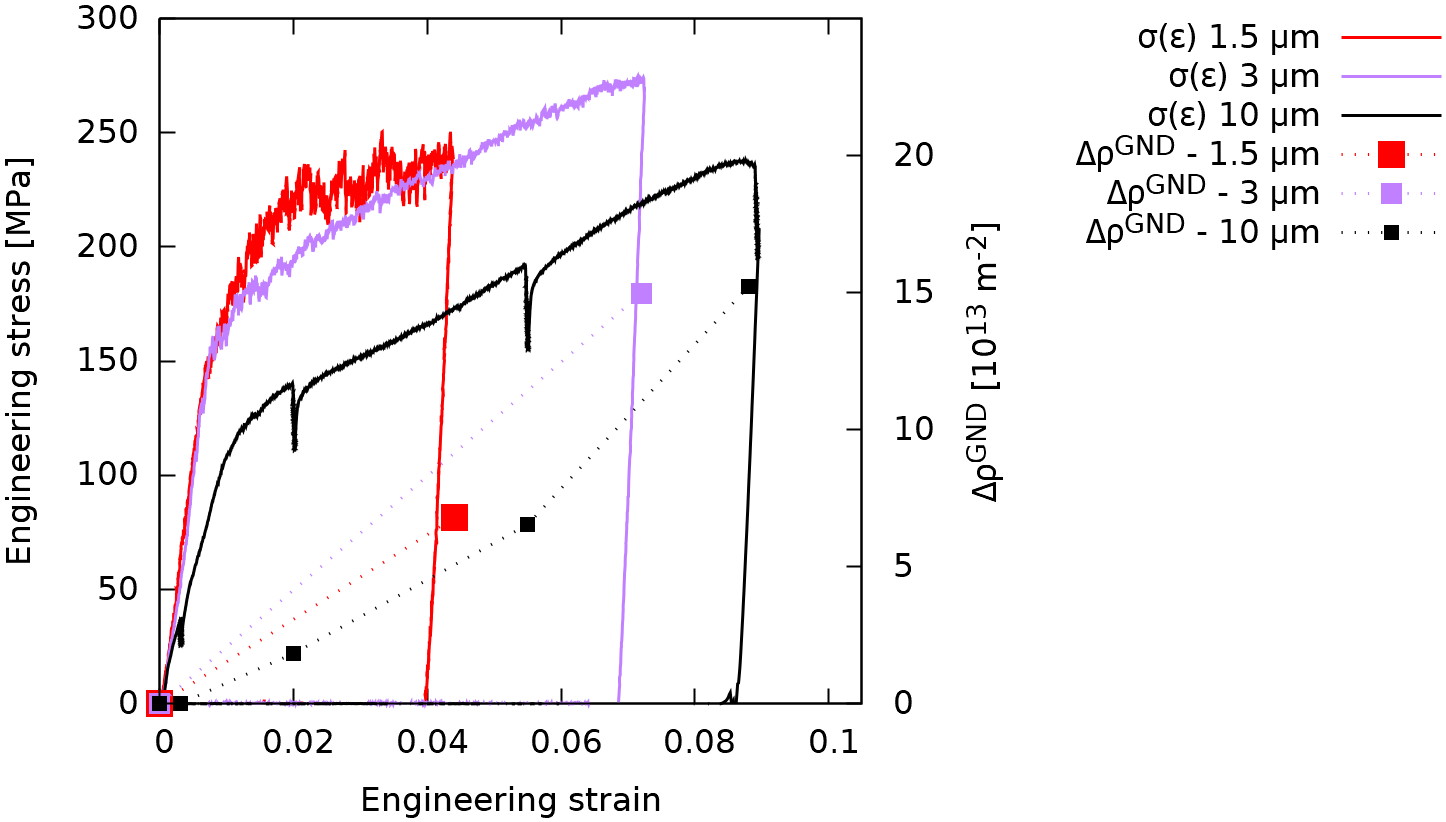}
    \hspace{0.1cm}
    \includegraphics[width=0.17\textwidth]{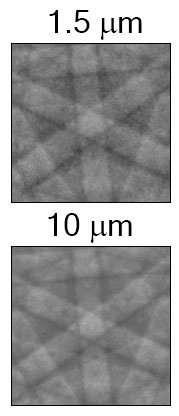}
    \caption{Left: Representative engineering stress-strain curves for different sized pillars (1.5-3-10 $\mu$m side length). Increase in the average GND density ($\Delta \rho^\mathrm{GND}$) is also plotted as a function of the engineering strain. The stress drops seen for the 10 $\mu$m pillar correspond to \emph{in situ} HR-EBSD measurements performed in order to get a more detailed picture on the GND density evolution. Right: Examples of the diffraction patterns of the $[1\bar{1}1]$ zone axis in case of the 1.5 $\mu$m and the 10 $\mu$m pillars.}
    \label{fig:sigma-epsilon}
\end{figure*}

Micropillars were deformed in situ to study how the GND density evolution happens at subsequent loading steps (Figure \ref{fig:sigma-epsilon}). During the experiment, the deformation process was paused while the HR-EBSD mapping was performed on the surface of the pillars. When the process is paused (for approx.~10--25 minutes during mapping) stress drops can be observed due to stress relaxation in the material and some minor creep of the piezoelectric actuator and the load cell (as seen in Figure \ref{fig:sigma-epsilon} $\sigma(\epsilon)$ curve of the 10 $\mu$m pillar). Other pillars were deformed from the initial to final stage without these intermittent mapping pauses (Figure \ref{fig:sigma-epsilon} $\sigma(\epsilon)$ curve of the 1.5 and 3 $\mu$m pillars). Size related hardening effect is also evident here, as not only the yielding point shifts to higher stresses at smaller pillar sizes, but the steepness of the $\sigma(\epsilon)$ in the plastic regime also decreases.

Based on the acquired stress-strain curves, it is evident that in smaller sized pillars larger strain bursts were detected, as the noisiest curve belongs to the 1.5 $\mu$m pillar, in contrast to the rather smooth $\sigma(\epsilon)$ curve of the 10 $\mu$m pillar. GND density evolution estimated by the $L^1$ optimization method is quite similar in all samples, showing a significant increase in $\Delta \rho^\mathrm{GND}$ upon loading. GND density tends to increase faster in smaller pillars, that is most likely an artifact due to the higher EBSD pattern noise level at smaller pillars. In case of reduced volumes, diffraction patterns have increased noise (see Figure \ref{fig:sigma-epsilon} right), that inherently increases the obtained values of $\rho^\mathrm{GND}$.

A series of identically prepared micropillar compression experiments were performed and the resulting $\sigma(\epsilon)$ curves are presented in Appendix A. Note, that the observations made above for the exemplary $\sigma(\epsilon)$ curves apply for all the measured ones. As expected at this size regime, the curves are of stochastic nature, that is, they differ from pillar to pillar due to the randomness of the initial internal microstructure. Interestingly, the highest scatter in the measured stress values do not seem to decrease with increasing size and are still significant at $a=10 \mu$m.

\begin{figure*}[!ht]
    \centering
    \includegraphics[width=0.99\textwidth]{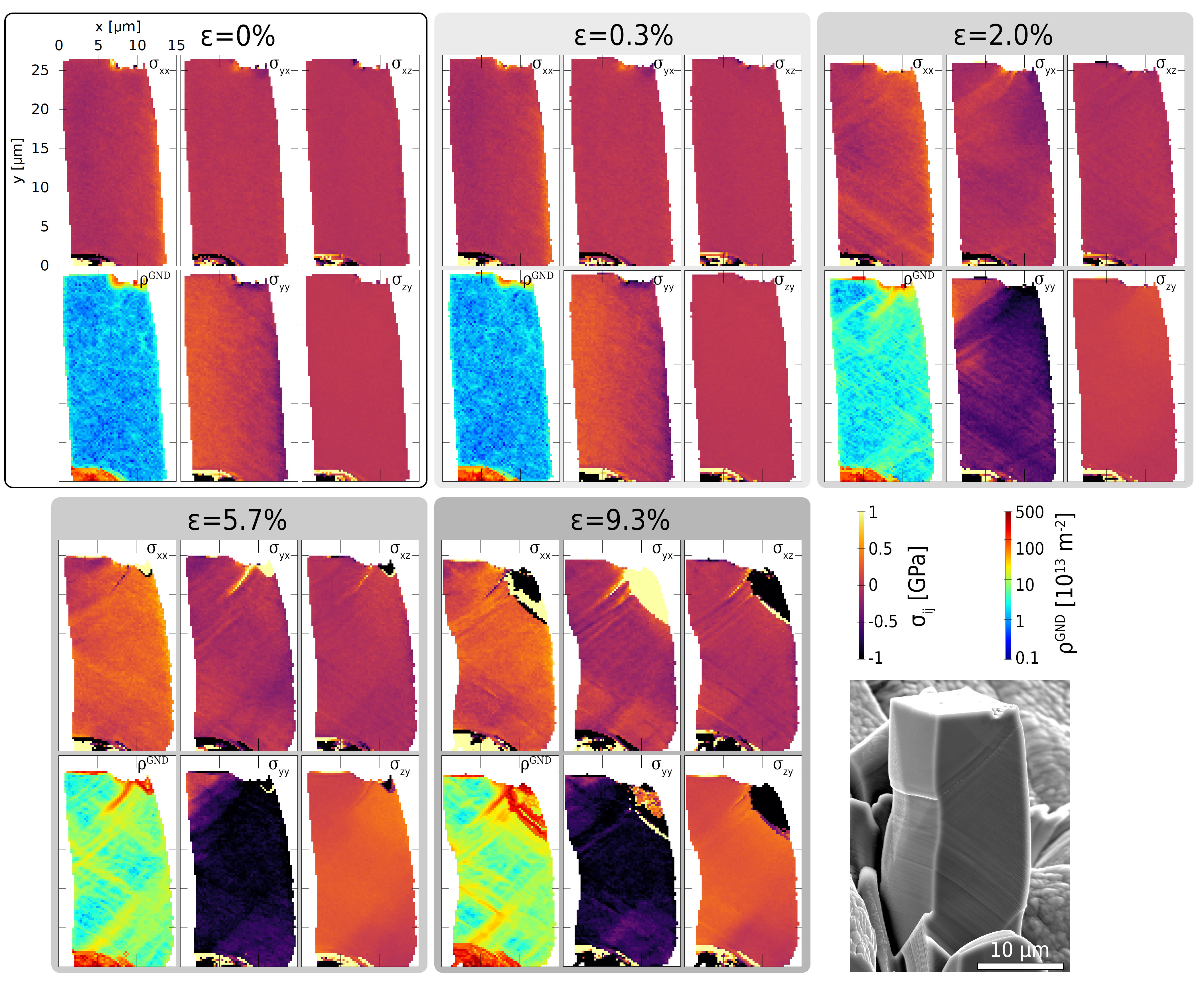}
    \caption{Stress tensor components of the 10 $\mu$m pillar deformed in situ. Corresponding $\sigma(\epsilon)$ curve is plotted in Figure \ref{fig:sigma-epsilon} (black line). Secondary electron image of the pillar after deformation is shown in the bottom right corner.}
    \label{fig:sigmacomp}
\end{figure*}

In situ HR-EBSD results are shown in Figure \ref{fig:sigmacomp} for the 10 $\mu$m pillar. The corresponding $\sigma(\epsilon)$ curve can be seen in Figure \ref{fig:sigma-epsilon} in black. It is worth mentioning that a flat punch tip with a diameter of 15 $\mu$m was chosen for the in situ compression in order to avoid shadowing effect on the measured side and maximize the usable surface for HR-EBSD evaluation. 
As it can be seen in the supplementary video recorded during the compression, a hard tiny particle attached to the punch tip caused a small impression on the top of the pillar, that acted as a source for dislocations to be created in higher numbers in that region. This however was not expected to influence the deformation in the lower parts of the micropillar.

The first map was taken prior to deformation, showing the initial stress state and dislocation density in the pillar. Components of the $\sigma_{ij}$ stress tensor in the 0\% case show no stress concentration in the sample, as expected. Please note the increased values at the top and bottom part of the pillar. These are only artifacts coming from the proximity of the tip and the FIB milling around the pillar. In the top part, the nearby tip acts as excess volume of material that electrons need to pass through before reaching the detector, that causes the diffraction pattern to decrease in quality close to the tip. At the bottom part, remaining of the FIB milled bulk material can act similarly, reducing the number of electrons that can reach the sample and detector surface. Decreasing the image quality can increase the inaccuracy of the cross-correlation based image analysis, hence causing the off-scale values seen at the sample extremities. One diffraction pattern was chosen from this $\epsilon=0\%$ map as reference (close to the base of the pillar), and all pixels were compared to this point, even at higher strains. This enabled that the $\sigma_{ij}$ values are related to the "stress-free" state of the material, making the evaluation unified throughout the experiment.

The second map was taken while the sample was loaded elastically. Naturally, the map taken at 0.3\% strain shows no major difference from the unloaded state. No notable GND density increase has been detected.

Continuing the compression process, the third map was taken at the onset of plastic deformation, at 2.0\% strain. Increased activity in all $\sigma_{ij}$ components was registered. As seen in the $\sigma_{yy}$ component, compressive stresses dominate the whole sample, while GND density increases significantly.

At higher strains, slip activity is noted in all stress component maps, while the average GND density reached its maximum value at the highest load of $\epsilon \approx$ 9\%. Looking at the surface of the pillar, slip bands clearly show multi-directional slip activation. Interestingly, when the GND density distribution is compared to the visible slip traces on the surface, a directional difference is observed. As slip traces on the surface of the pillar correspond to dislocations that already left the system, the GNDs accumulating inside the material do not necessarily align with these slip planes. Indeed, it can be concluded that in some cases GNDs seem to pile up in directions perpendicular to the visible surface slip bands.

\subsection*{Burgers vector analysis}

\begin{figure*}[!ht]
    \centering
    \includegraphics[width=0.9\textwidth]{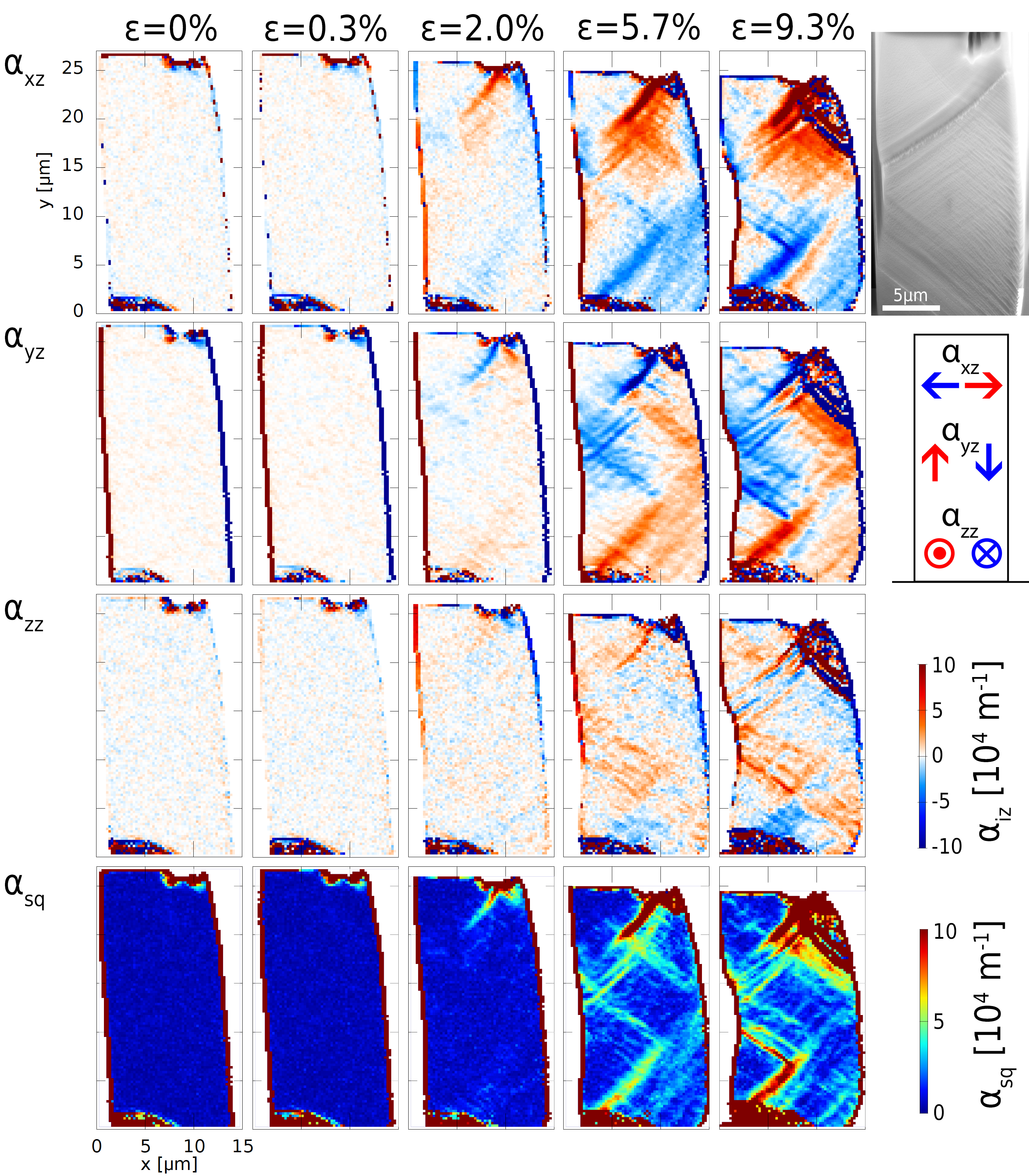}
    \caption{The experimentally accessible components of the Nye tensor of the 10 $\mu$m pillar deformed in situ. Corresponding $\sigma(\epsilon)$ curve is plotted in Figure \ref{fig:sigma-epsilon} (black line). A sketch of the Burgers vector directions is also included on the right-hand-side, indicating that the sign of the $\alpha_{iz}$ components can be linked to specific types of GNDs. Secondary electron image shows the slip planes on the studied surface.}
    \label{fig:alphacomp}
\end{figure*}

Figure \ref{fig:alphacomp} plots the accessible $\alpha_{iz}$ components of the Nye tensor for all studied strains. Similarly to the stress tensor components, the first two stages show no remarkable features in any of the $\vec \alpha$ maps. Increased dislocation activity is only evident at higher strains in the plastic regime, where a rich GND structure is seen to develop with characteristic features in all three tensor components.

According to the definition of the Nye dislocation density tensor in equation (\ref{eq:01}), the $\alpha_{iz}$ components characterize the average Burgers vector of dislocations being parallel to the $z$ axis, that is, perpendicular to the sample surface. In particular, $\alpha_{xz}$ and $\alpha_{yz}$ components in Figure \ref{fig:alphacomp} represent edge components of the GNDs, while $\alpha_{zz}$ represents a pure screw component as shown by the complementary sketch in Figure \ref{fig:alphacomp}.

The vector constructed from the three available Nye tensor components $\vec B = (\alpha_{xz}, \alpha_{yz}, \alpha_{zz})^T$, is, therefore, parallel to the average Burgers vector of dislocations parallel to the $z$ axis. Since at a given measurement point dislocations on various slip systems may be simultaneously present, one expects $\vec B$ to be a linear combination of different Burgers vectors with weight factors representing the fraction of dislocations in given slip systems. If, however, $\vec B$ happens to be parallel with one of the Burgers vectors of the crystal one can conclude, that GNDs are dominated by dislocations with that specific Burgers vector.

\begin{figure*}[!ht]
    \centering
    \includegraphics[width=0.9\textwidth]{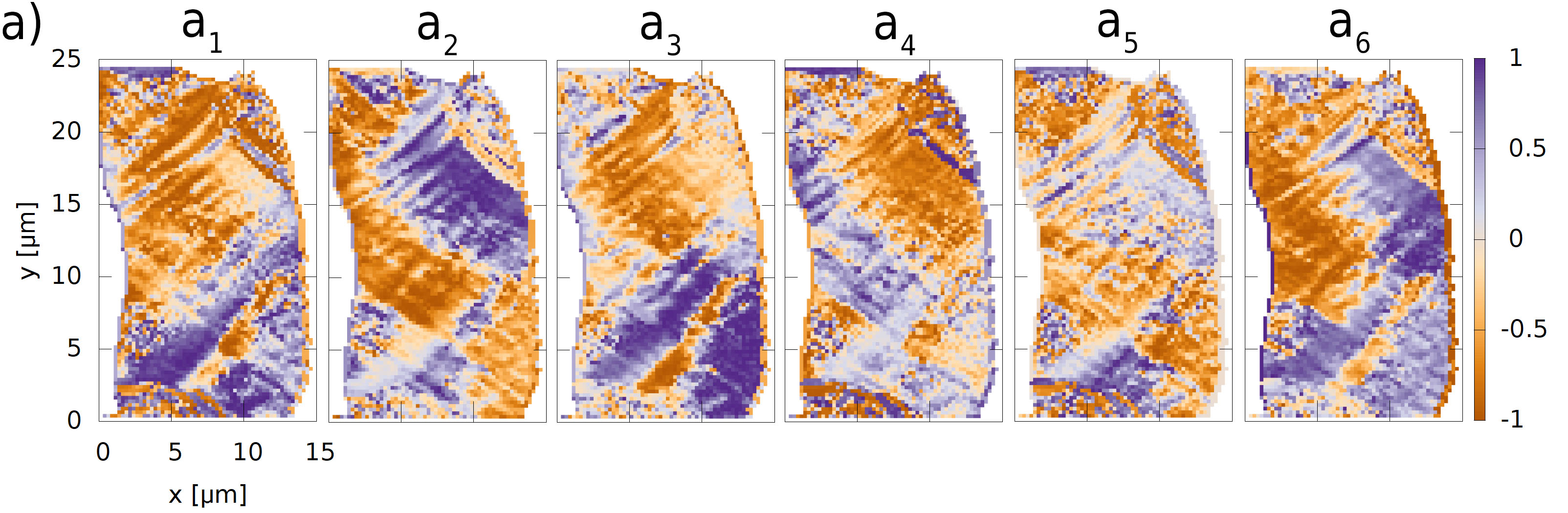}
    \hspace{0.1cm}
    \includegraphics[width=0.5\textwidth]{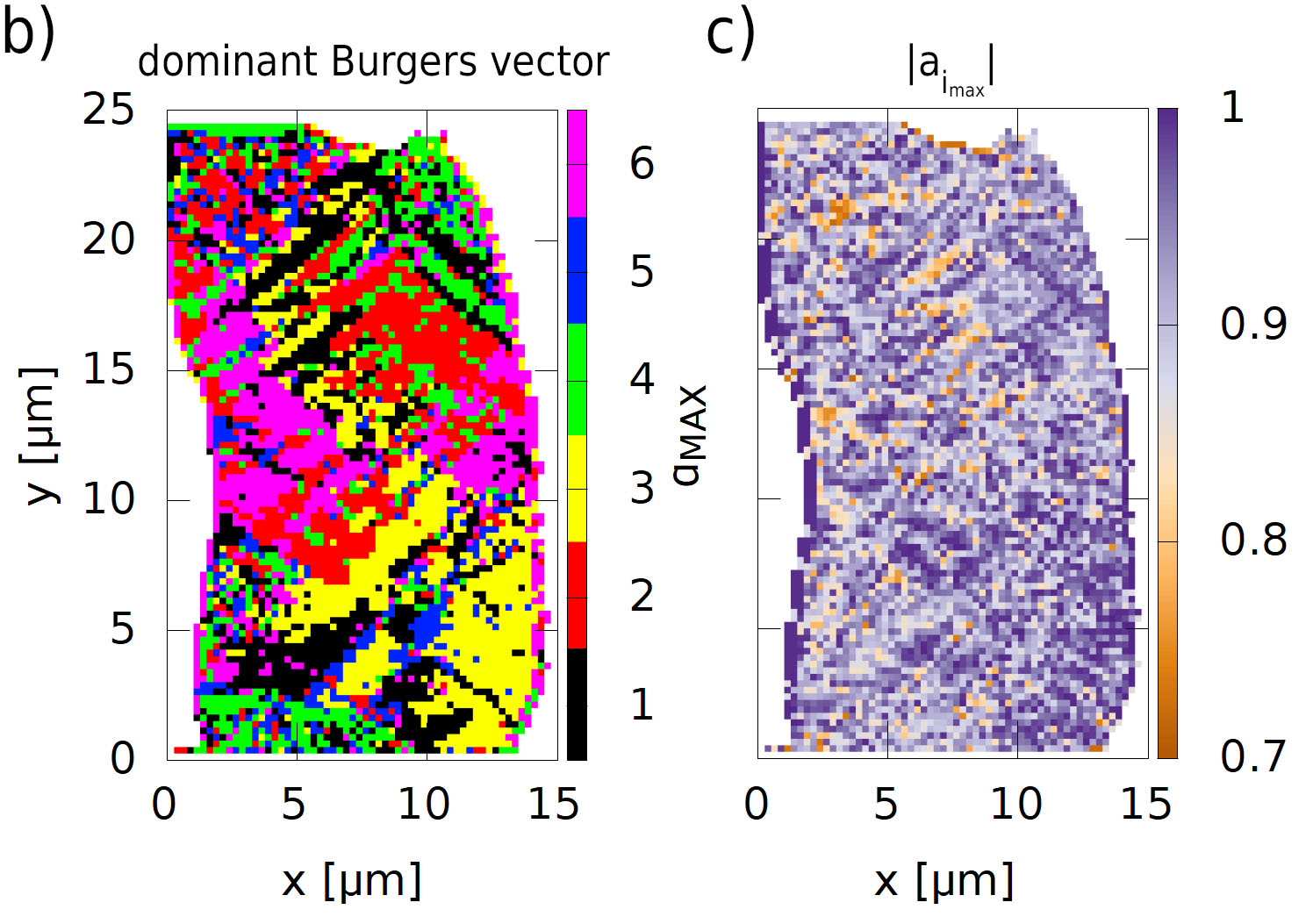}
    \caption{(a) $a_i$ values (defined by equation (\ref{eq:14})), (b) dominant Burgers vectors and (c) $|a_{i_\mathrm{max}}|$ values plotted for the 10 $\mu$m pillar at $\epsilon=9.3\%$ deformation.}
    \label{fig:acomp}
\end{figure*}

In this spirit, we define the values \begin{eqnarray}
a_i = \frac{\vec B \cdot \vec b_i}{B b_i}\quad i=1,\dots,6, \label{eq:14}
\end{eqnarray}
where $\vec b_i$ is the Burgers vector corresponding to slip direction $i$ (see Figure \ref{fig:2a_system}) expressed in the $xyz$ coordinate system. The values $a_i$ are, thus, equal to $\cos(\varphi_i)$, with $\varphi_i$ being the angle between $\vec B$ and $\vec b_i$. In Figure \ref{fig:acomp}(a) plots of the $a_i$ are seen and values of $\approx \pm1$ ($\approx 0$) mark regions where $\vec B$ is close to parallel with (perpendicular to) the given Burgers vector.


It is evident from Figure \ref{fig:acomp} that there are extensive regions with $|a_i| \approx 1$ with various $i$ values. This suggests, that these regions are characterized by the dominance of one type of Burgers vector. In order to better visualize this finding, Figure \ref{fig:acomp}(b) plots at every measurement point the Burgers vector index $i_\mathrm{max}$ that has the highest $|a_i|$ value there, and Figure \ref{fig:acomp}(c) shows the $|a_{i_\mathrm{max}}|$ values. From the latter one can conclude that in $\sim 90$\% of the sample surface the angle between $\vec B$ and $\vec b_{i_\mathrm{max}}$ is less than 27$^\circ$ and in $\sim 35$\% less than 18$^\circ$. So, as expected, $\vec B$ is not exactly parallel with one of the possible Burgers vectors (as it represents an average Burgers vector in the volume element measured by HR-EBSD) but one can clearly find a dominant Burgers vector in most regions of the sample. According to Figure \ref{fig:acomp}(b) regions with identical dominant Burgers vector form a granular-like anisotropic structure with preferred directions parallel with the possible slip directions of the $B$ and $C$ slip planes (see Figure \ref{fig:2a_system}). We recall, that the active slip planes are ${B2, B4, C1, C3}$ where most of the plastic slip takes place. Interestingly, regions with dominant  Burgers vector type 1 and 3 are mostly aligned along the $B$ plane and vice versa, those with Burgers vector type 2 and 4 mostly along the $C$ plane. This suggests that the GNDs are not accumulated on the active slip systems, rather on inactive ones. The fact that the GNDs are yet aligned parallel to an active slip plane is probably because these GNDs were generated by dislocation reactions or cross slip of dislocations moving in the active slip system. In addition, significant regions with Burgers vectors of type 6 are found which is of pure edge character and do not belong to any of the active slip systems.

\subsubsection*{Size dependence of the GND density network}

\begin{figure*}[!ht]
    \centering
    \includegraphics[width=0.9\textwidth]{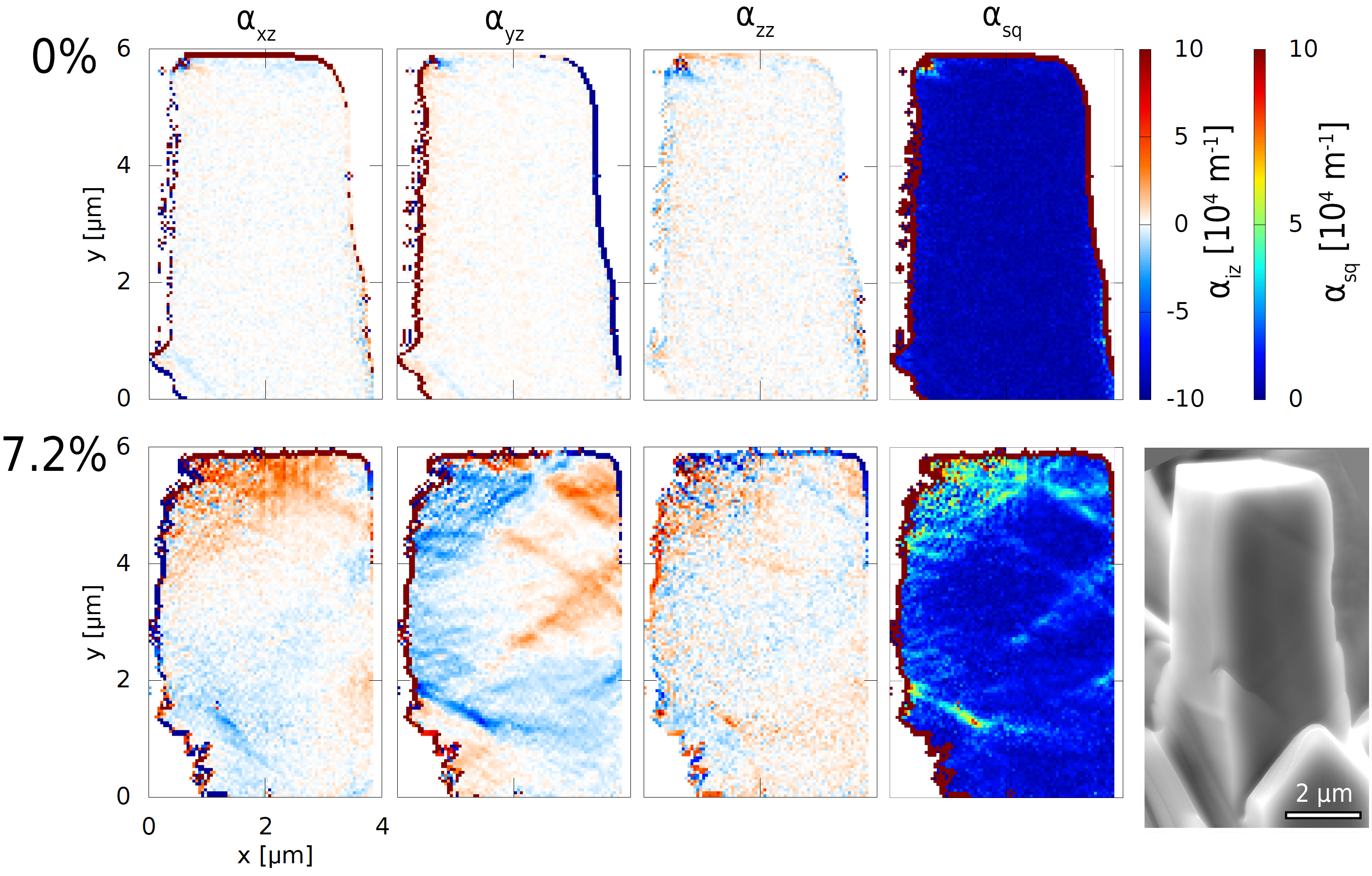}
    \caption{Nye tensor components of the 3 $\mu$m pillar at $\epsilon=0\%$ and $7.2\%$ deformation. Corresponding $\sigma(\epsilon)$ curve is plotted in Figure \ref{fig:sigma-epsilon} (purple line). A sketch of the Burgers vector directions is shown in Figure \ref{fig:alphacomp}.}
    \label{fig:3um}
\end{figure*}

As already mentioned in the introduction, the GND density network strongly depends on the size of the pillar. In case of the 10 $\mu$m pillar, GNDs in the middle of the sample form a cellular dislocation structure (see Figure \ref{fig:alphacomp}). If the size of the specimen is reduced, the presence of the surface affects even more the collective behaviour of the dislocations. As it is shown in Figure \ref{fig:3um}, at the diameter of 3 $\mu$m the formed GND network after $\sim 7\%$ deformation looks quite different from the already presented 10 $\mu$m specimen. This pillar was only measured \emph{post mortem}, after the deformation process had finished and the sample was no longer under load. Although some slip planes can be identified, the cellular structure is completely missing. This is a sign that as the pillar size approaches the 1 $\mu$m limit reported in the literature, a transition from dislocation accumulation to dislocation starvation takes place. If the surface of the specimen is examined, slip planes are still clearly observed but GNDs are stored in significantly less amount compared to the 10 $\mu$m pillar as evident from the $\alpha_\mathrm{sq}$ map in Figure \ref{fig:3um}.

\begin{figure*}
    \centering
    \includegraphics[width=0.9\textwidth]{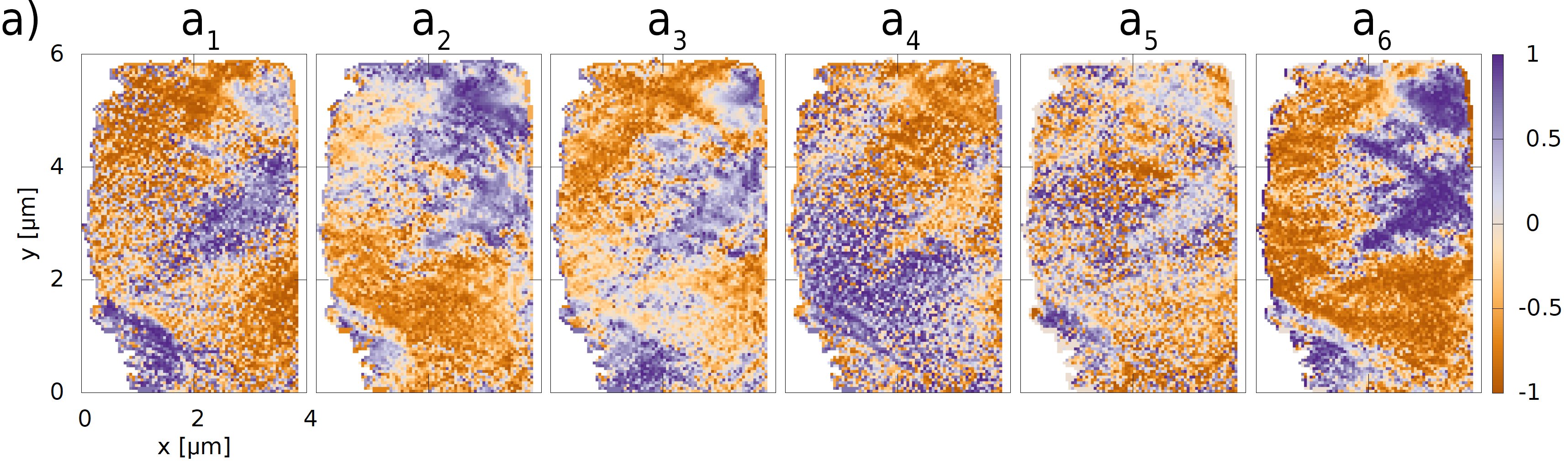}
    \hspace{0.1cm}
    \includegraphics[width=0.5\textwidth]{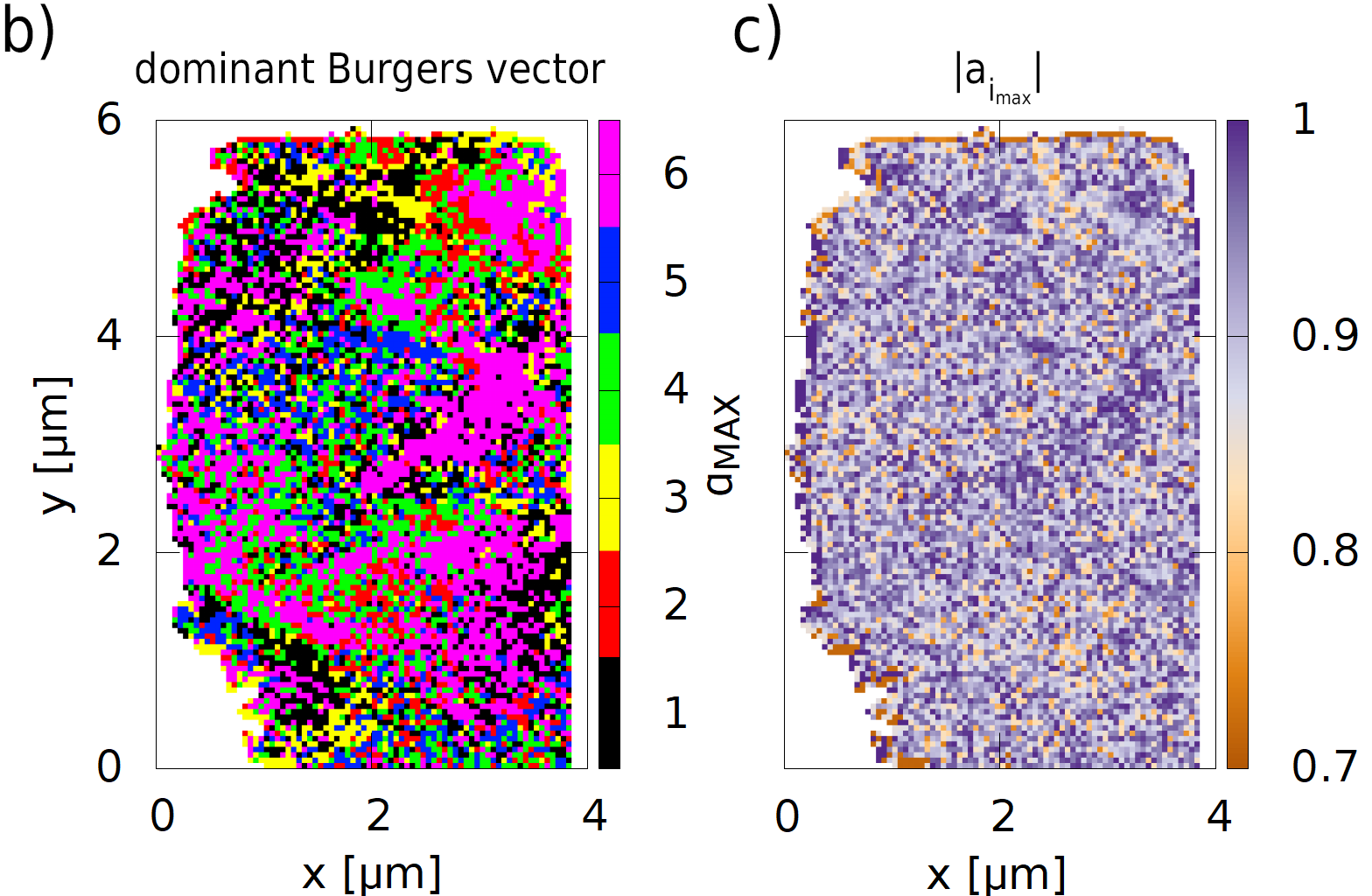}
    \caption{(a) $a_i$ values, (b) dominant Burgers vectors and (c) $|a_{i_\mathrm{max}}|$ values plotted for the 3 $\mu$m pillar at $\epsilon=7.2\%$ deformation.}
    \label{fig:3um-a}
\end{figure*}

In Figure \ref{fig:3um-a} (a) extensive regions with $|a_i| \approx 1$ can be identified. From the Burgers vector plots in Figure \ref{fig:3um-a} (b) one mostly cannot find clear regions that are characterized by the dominance of only one type of Burgers vector, unlike in the larger pillar. At best, some regions with Burgers vector type 6 are observed, but this (similarly to the 10 $\mu$m pillar) cannot be linked to any of the active slip systems. Figure \ref{fig:3um-a} (c) enforces our conclusion that a dominant Burgers vector cannot clearly be identified in this specimen.

\subsection{Simulations}
\label{ssec:3c_Simulations}

The fcc copper micropillars are modelled by the CDD formulation presented in section \ref{ssec:2c_Simulations}. 
The elastic constants are given as $C_{1111}=168\,\mathrm{GPa}$, $C_{1122}=121\,\mathrm{GPa}$ and $C_{2323}=75\,\mathrm{GPa}$ \cite{ledbetter.1974,rosler.2019} leading to a Zener anisotropy of $A=3.22$ \cite{ledbetter.1974,rosler.2019}. 
The isotropic elastic constants needed for the internal stress calculation are determined following \cite{date.1969} as: 
$G=40.0\,\mathrm{GPa}$, $\nu=0.367$. 
The length of the Burgers vector is given as $b=0.254\,\mathrm{nm}$. 
The material specific drag coefficient is $B=5\times10^{-5}\,\mathrm{sPa}$ \cite{kubin.1992}.
The coefficients of the material interaction matrix used in the yield stress term are chosen according to \cite{kubin.2008} as: $a_\mathrm{self}=0.122,\, a_\mathrm{Hirth}=0.07,\, a_\mathrm{Lomer}=0.122,\, a_\mathrm{gliss}=0.137,\, a_\mathrm{coll}=0.625$.
The backstress parameter is chosen as derived in \cite{schmitt.2015} as $D=0.256$.
The 
constant for the cross-slip probability term is assumed as $\beta=10^5$ according to  \cite{weygand.2002}, the activation volume is $V_\mathrm{act}=300\,b^3$ \cite{bonneville.1988} and the stress initializing stage-three hardening is set to $\tau_\mathrm{III}=0.028\,\mathrm{GPa}$ \cite{kubin.1992}. 
The collision rate constants for the dislocation reactions, cp. \cite{sudmanns.2020}, are set to $c_\mathrm{gliss}=2c_\mathrm{coll}=0.64$ and $c_\mathrm{Lomer}=640$. 

We study specimens with a $\langle1\,1\,0\rangle$ crystal orientation and with a geometric dimension of $a \in \{3,6,10\}\,\mu\mathrm{m}$ according to Figure \ref{fig:2a_system}.
The system is subjected to a displacement at the top of the system in axial direction ($\vec u_\dir |_{y=3a} = -u_y\,\vec e_y$) with a rate of $\dot{\varepsilon}_{yy}=-4\times10^3\,\nicefrac{1}{\mathrm{s}}$ 
up to a maximum loading of $\varepsilon_{yy}=-10\,\%$ . The bottom of the system is fixed ($\vec u_\dir|_{y=0} = \vec 0$), the circumference surfaces are considered traction free. The critical resolved shear stress of $\tau^\mathrm{CRSS} \in \{101,65,47\}\,\mathrm{MPa}$ is given according to \cite{wu.2016}.
An initial total dislocation density is given as $2.4 \times 10^{12}\,\nicefrac{1}{\mathrm{m}^2}$ equally allocated to the different slip systems and homogeneously distributed over the specimen. The initial dislocation density is assumed to consist purely of dislocation network density, which consist half of stabilized dislocation density and half of Lomer junctions. 
The considered numerical system resolution is given by a mesh of approximately $12.3 \times 10^3$ elements leading to an element width of $\nicefrac{a}{16}$ and a total number of degrees of freedom of about $10^6$.
Taking the Courant-Friedrichs-Lewy (CFL) condition into account, a smaller time step have had to be used for the micropillar of $a =3\,\mu m$ size. Therefore, the maximum applied load was reduced in this case.

\begin{figure*}[!ht]
    \centering
    \includegraphics[width=0.4\textwidth]{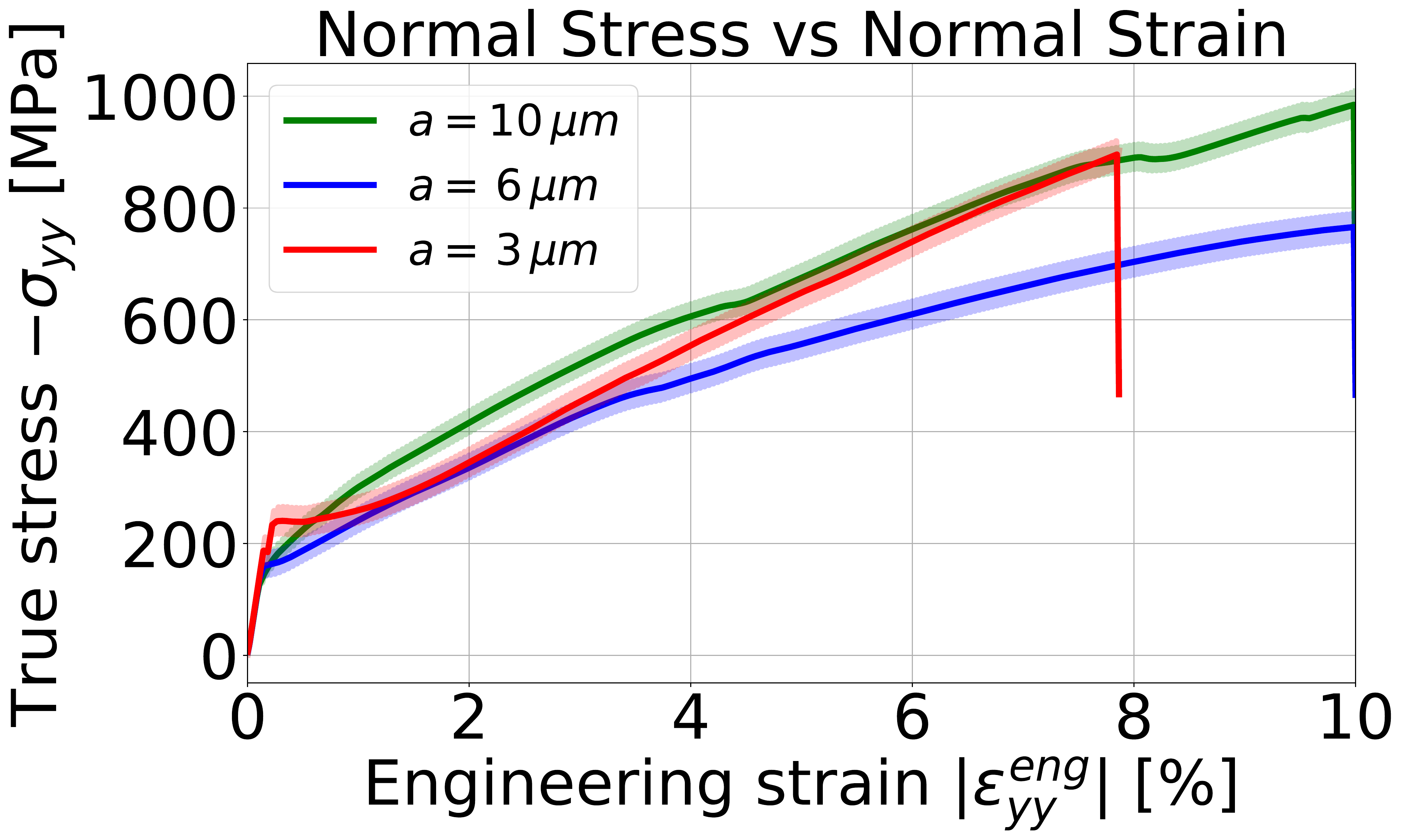}
    \hspace{1cm}
    \includegraphics[width=0.375\textwidth]{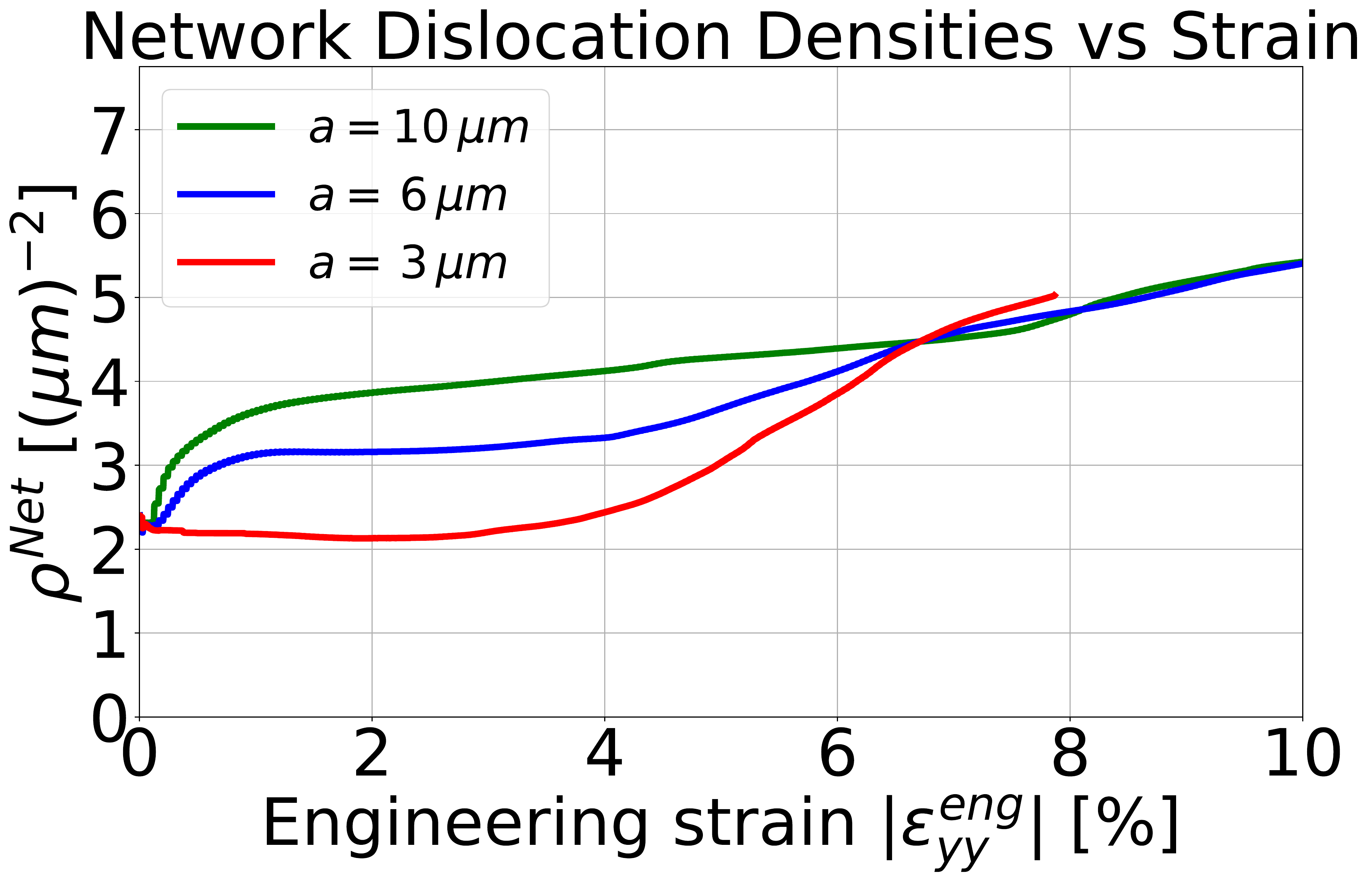}
    \caption{Stress-strain curves (left) and the evolution of the 
    network dislocation density (right) for varying micropillar sizes. 
    }
    \label{fig:3c_size_sig}
\end{figure*}

\subsubsection*{Size dependency}

To investigate the influence of the micropillar size on the formation of dislocation networks, in the following we vary the micropillar size.
The corresponding stress-strain curves and the evolution of dislocation network density are shown in Figure \ref{fig:3c_size_sig}.
Considering the micropillar sizes of $a \in \{3,6,10\}\,\mu\mathrm{m}$, the results of smaller pillars show an elastic-plastic transition at higher stresses with a sharper transition and lower hardening at the beginning of plastic yielding. 
For the chosen $\langle1\,1\,0\rangle$ crystal orientation the elastic-plastic transition occurs at approximately  $|\sigma_{yy}|\approx 235\,\mathrm{MPa}$ for $a=3\,\mu\mathrm{m}$ and $|\sigma_{yy}|\approx 125\,\mathrm{MPa}$ for $a=10\,\mu\mathrm{m}$.
Figure \ref{fig:3c_size_sig}(right) shows that the change of the amount of network dislocation density, which is pinned within the system, decreases for smaller micropillars. 
The microstructure evolution within the first two percent nominal strain leads to an increase of the network dislocation density of $60\,\%$ for $a=10\,\mu m$, of about $30\,\%$ for $a=6\,\mu m$ and a slight decrease of about $10\,\%$ for $a=3\,\mu m$.

\subsubsection*{Microstructure evolution}

\begin{figure*}[!ht]
    \centering
    \includegraphics[width=0.49\textwidth]{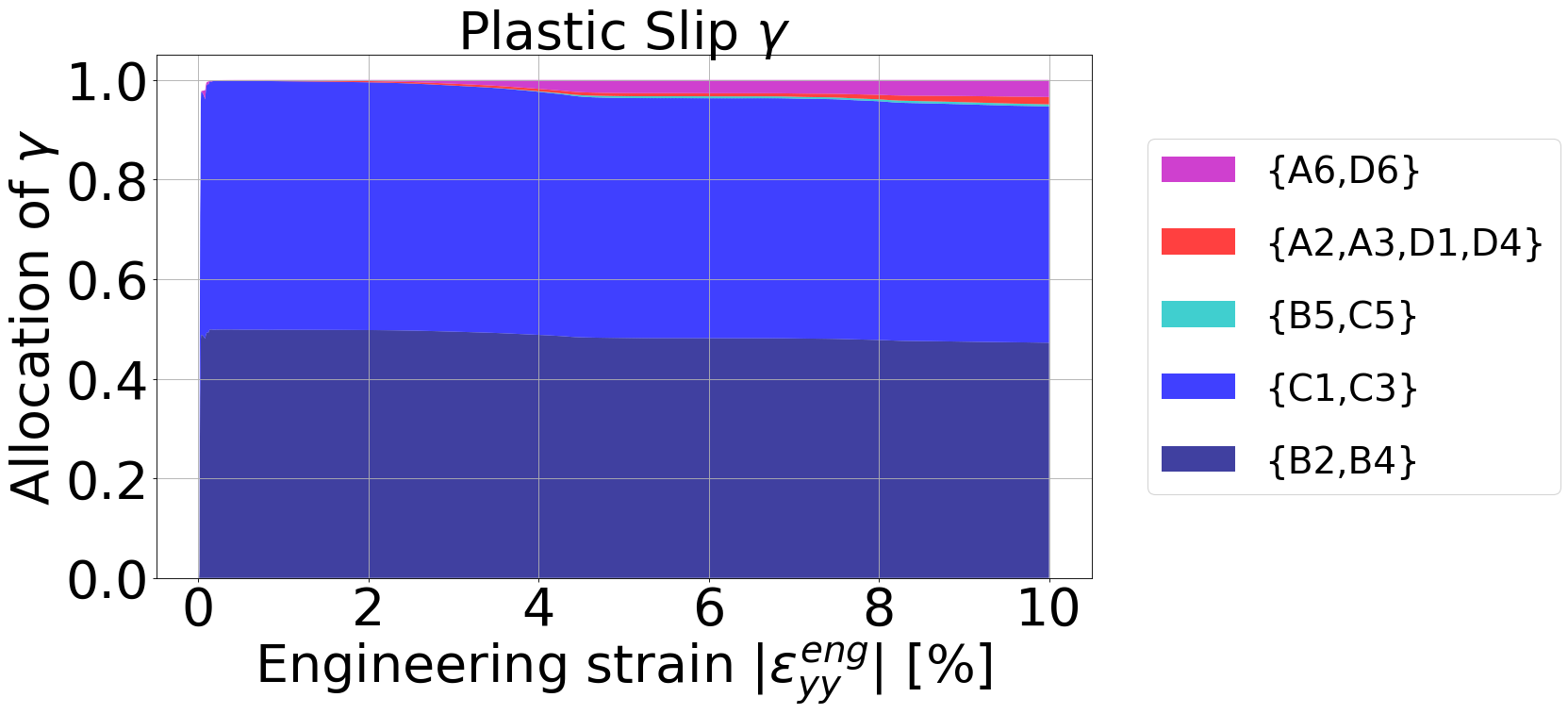}
    \hspace{0.01cm}
    \includegraphics[width=0.49\textwidth]{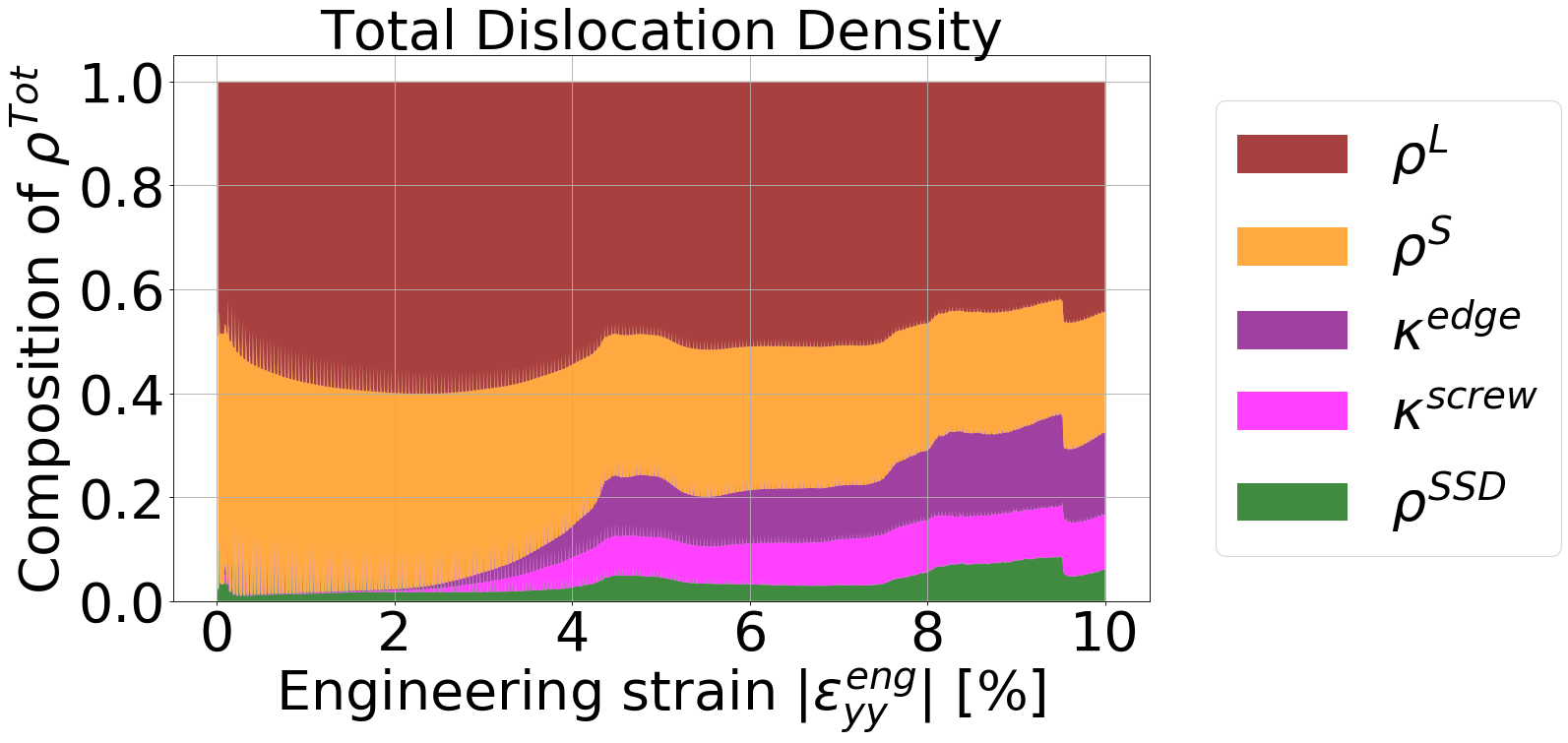}
    \caption{Evolution of the allocation of the plastic slip and the composition of the total dislocation density during the compression.}
    \label{fig:3c_struc_slip}
\end{figure*}

\begin{figure*}[!ht]
     \centering
     \begin{subfigure}[b]{0.3245\textwidth}
         \centering
         \includegraphics[width=\textwidth]{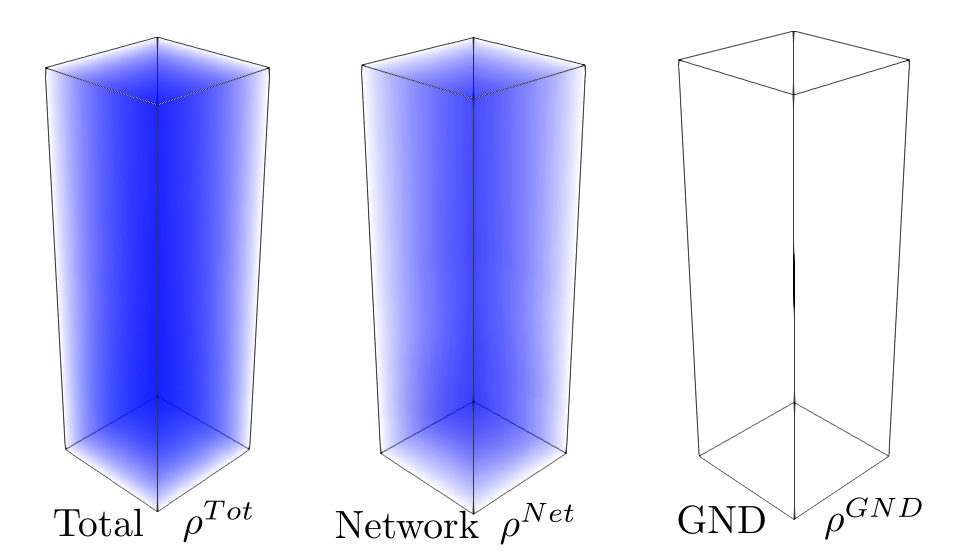}
         \caption{$\varepsilon_{yy}\approx-0.0\%$}
     \end{subfigure}
     \hfill
     \begin{subfigure}[b]{0.3245\textwidth}
         \centering
         \includegraphics[width=\textwidth]{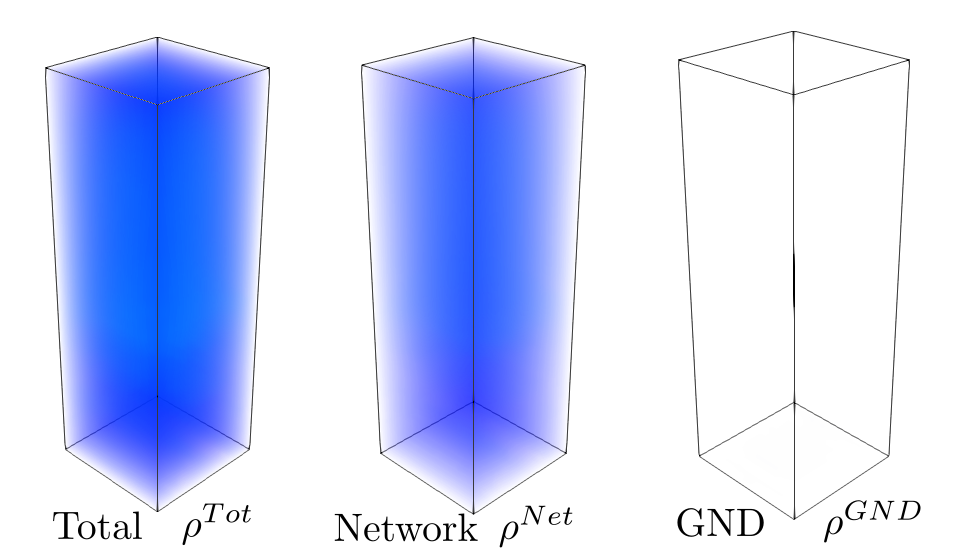}
         \caption{$\varepsilon_{yy}=-0.3\%$}
     \end{subfigure}
     \hfill
     \begin{subfigure}[b]{0.3245\textwidth}
         \centering
         \includegraphics[width=\textwidth]{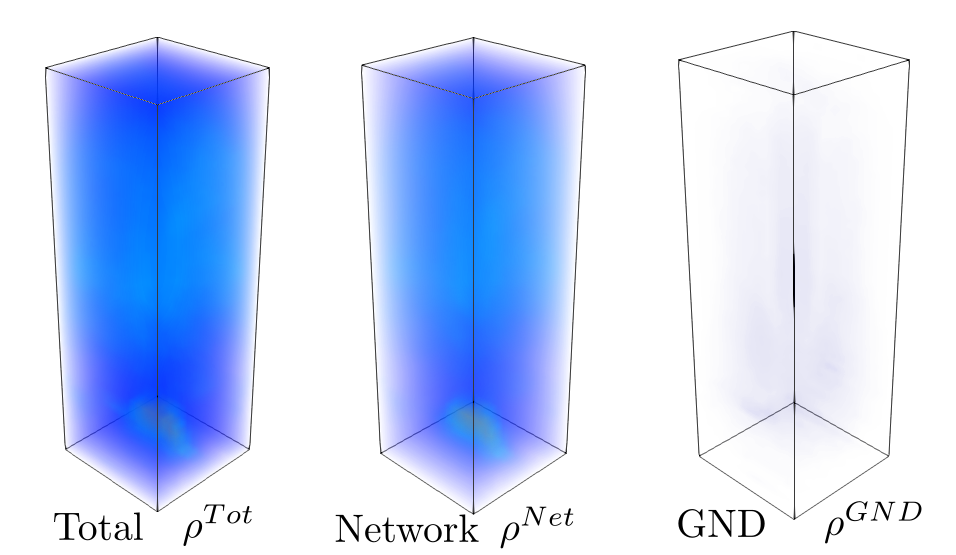}
         \caption{$\varepsilon_{yy}=-2.0\%$}
     \end{subfigure}
     \hfill
     \begin{subfigure}[b]{0.3245\textwidth}
         \centering
         \includegraphics[width=\textwidth]{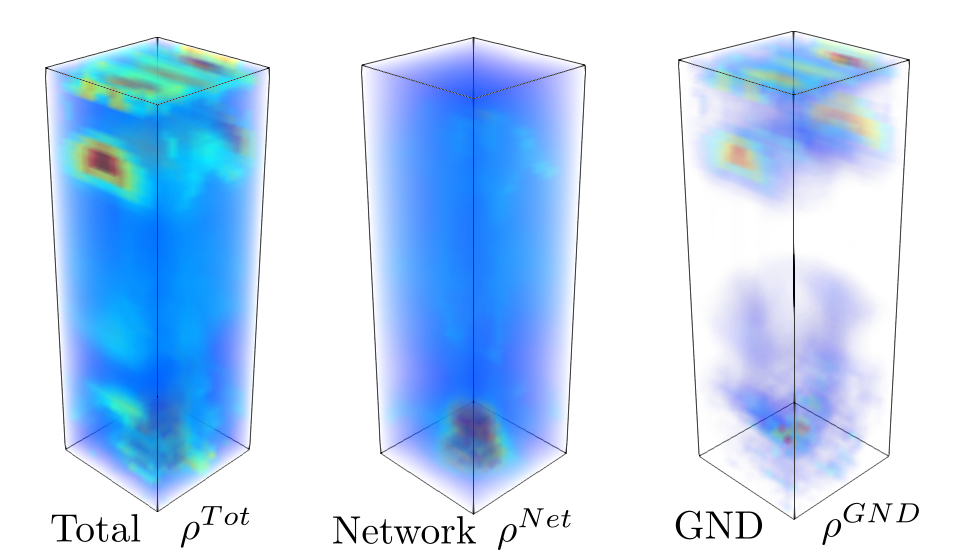}
         \caption{$\varepsilon_{yy}=-5.7\%$}
     \end{subfigure}
     \hspace{1cm}
     \begin{subfigure}[b]{0.3245\textwidth}
         \centering
         \includegraphics[width=\textwidth]{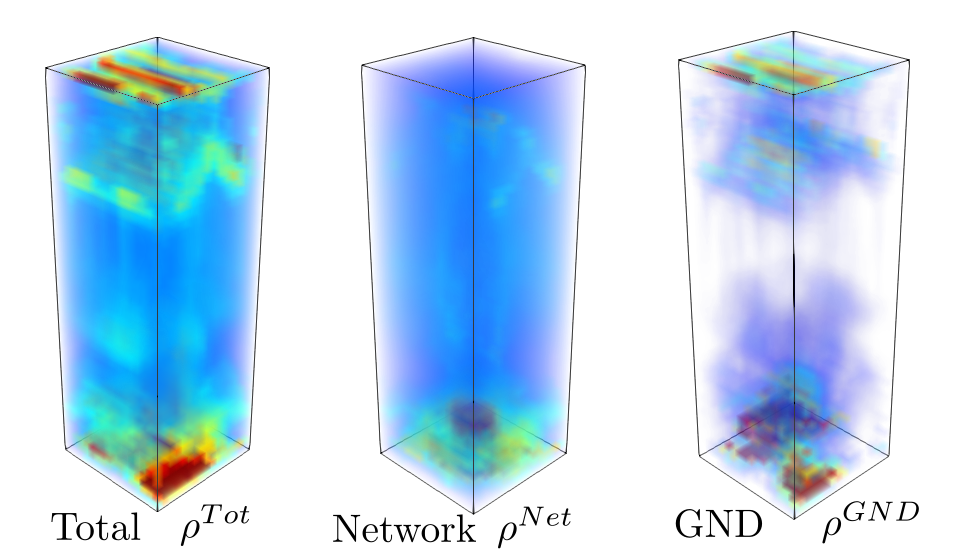}
         \caption{$\varepsilon_{yy}=-9.3\%$}
     \end{subfigure}
     \hspace{1cm}
     \begin{subfigure}[b]{0.1\textwidth}
         \centering
         \includegraphics[width=0.65\textwidth]{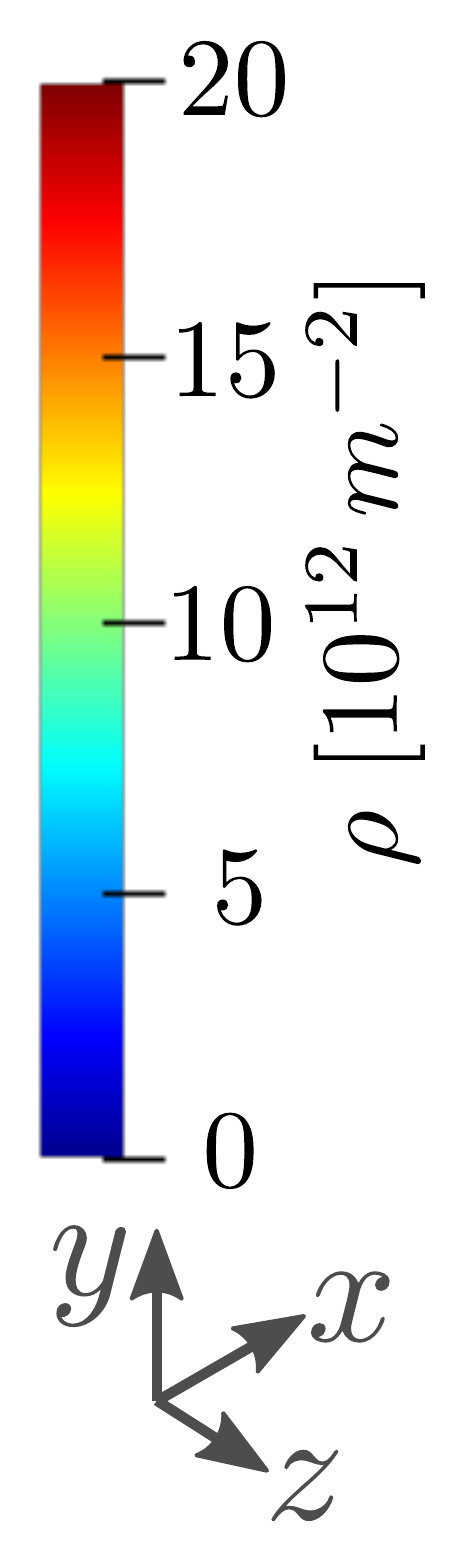}
     \end{subfigure}
        \caption{Spatial distribution of dislocation densities: total dislocation density, network dislocation density, and GND density shown for different strain states.
        }
        \label{fig:3c_struc_rho-vtk}
\end{figure*}

To gain deeper insights into the dislocation microstructure evolution, the evolution is exemplarily shown for the micropillar of size $a=10\,\mu\mathrm{m}$.
Figure \ref{fig:3c_struc_slip}(left) shows the plastic slip evolution over the strain considering its allocation to different slip systems. It can be observed that the plastic deformation mainly takes place on the four slip systems showing the highest Schmid factors, i.e.\ $\{B2,B4,C1,C3\}$, as shown in Figure \ref{fig:3c_struc_slip}(left). 
Finally, $95\,\%$ of the plastic deformation is caused by the activity of these slip systems.
Figure \ref{fig:3c_struc_slip}(right) depicts the total dislocation density during loading as composition of the different dislocation density quantities considered in the formulation. During the deformation, GND density is formed and maintained in the system. GND density represents a fraction of $26\,\%$ of the total dislocation density with respect to the considered system resolution by the end of loading. The fraction of SSD density of the total dislocation density is $6\,\%$ and thus lower than the contribution of GNDs. The remaining fraction of $68\,\%$ of the total dislocation density consists of network dislocation density.
The spatial distribution of different dislocation density quantities at exemplarily selected loading steps is shown in Figure \ref{fig:3c_struc_rho-vtk}.
%
%

\begin{figure*}[!ht]
    \centering
    \includegraphics[width=0.32\textwidth]{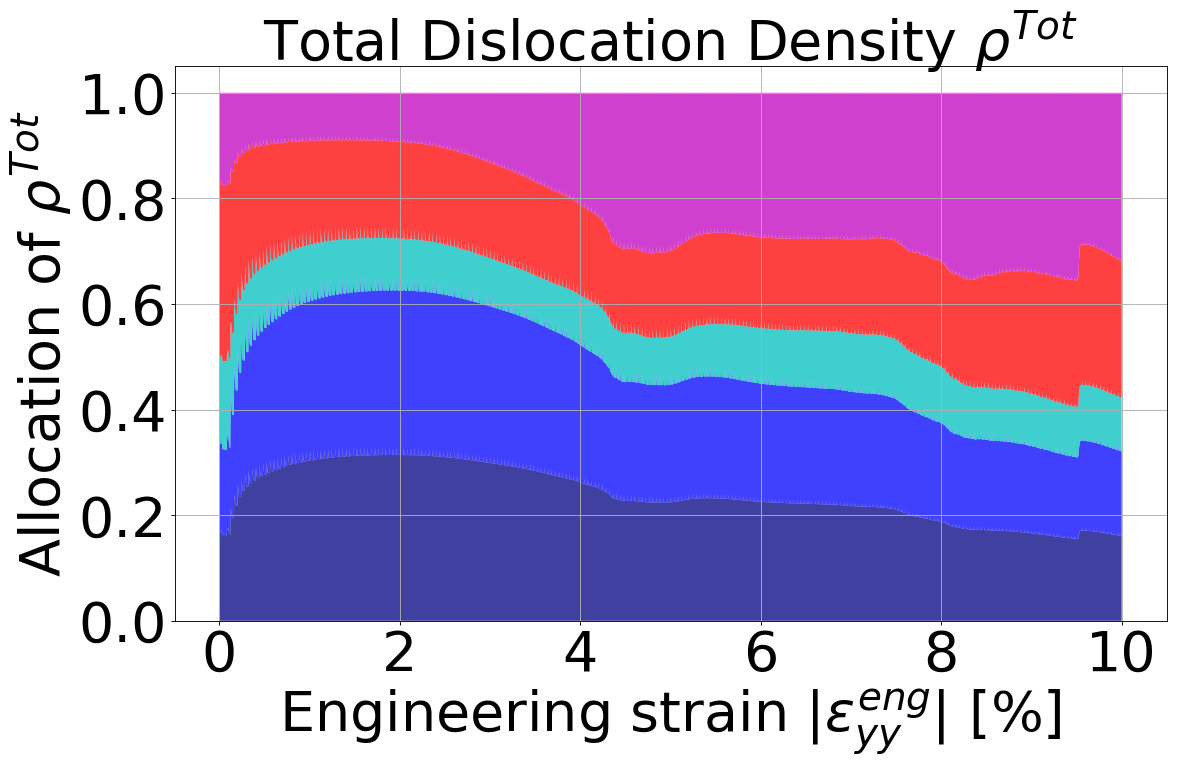}
    \hspace{0.01cm}
    \includegraphics[width=0.32\textwidth]{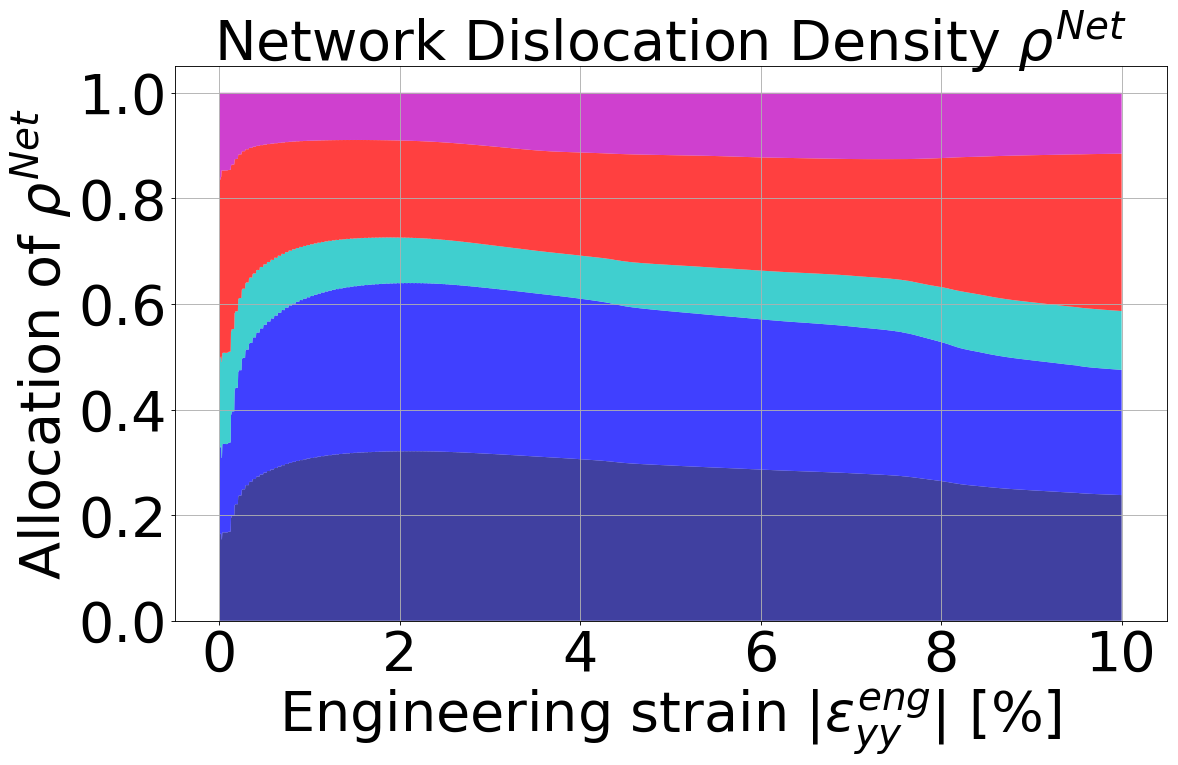}
    \hspace{0.01cm}
    \includegraphics[width=0.32\textwidth]{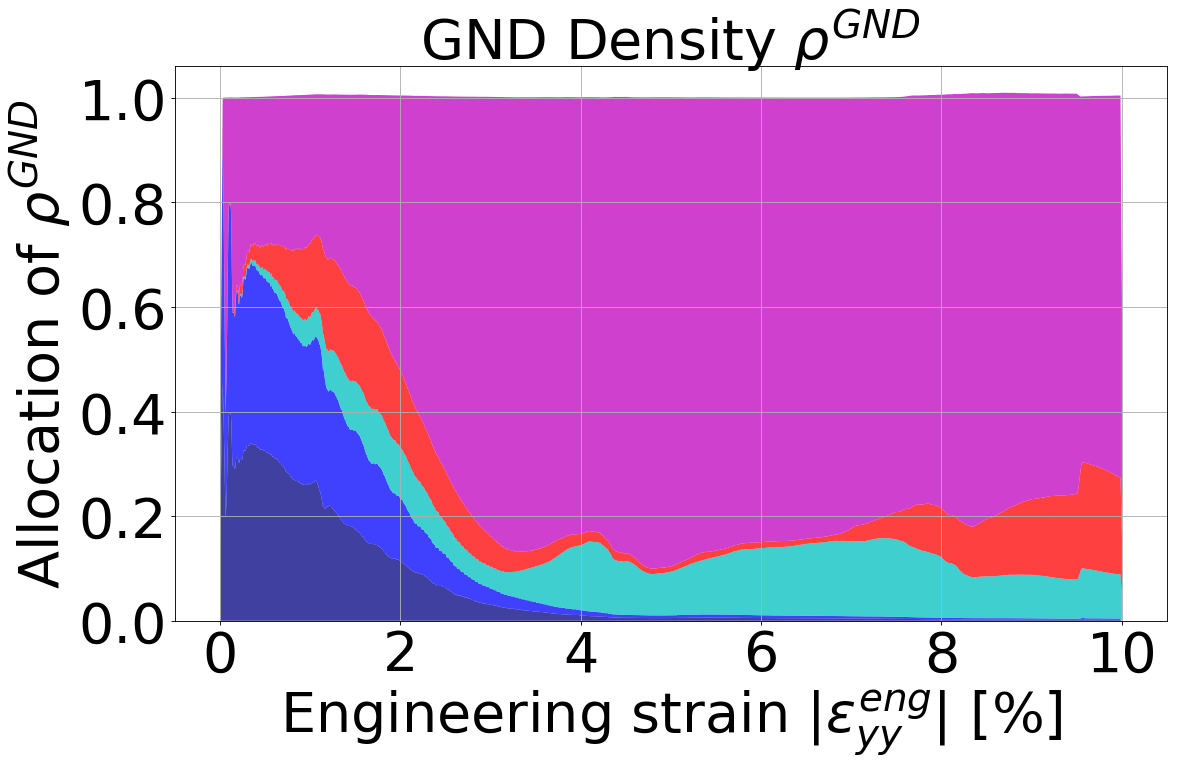}
    \hspace{0.01cm}
    \hfill
    \includegraphics[width=0.85\textwidth]{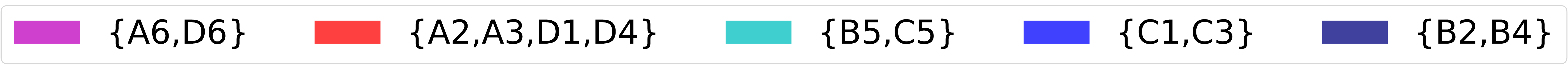}
    \hfill
    \caption{Allocation of the dislocation densities to the individual slip systems during the microstructure evolution.}
    \label{fig:3c_struc_alloc}
\end{figure*}

The contribution of the individual slip systems to the dislocation densities is shown in Figure \ref{fig:3c_struc_alloc}.
It can be observed that $32\,\%$ of the total dislocation density is located on the four active slip systems $\{B2,B4,C1,C3\}$ for $10\%$ strain. The remaining $68\,\%$ of the total density are on the eight inactive slip systems. However, differences can be observed in the allocation of density with respect to different types of dislocation density. For example, about half of the network density ($48\,\%$) is located on the four active slip systems, while the GND density is almost exclusively on the inactive slip systems. A significant amount of GND density ($73\,\%$) is located on the slip systems \{A6,D6\}, that have both slip plane normals and Burgers vector directions different from the active slip systems. 

\begin{figure*}[!ht]
    \centering
    \includegraphics[width=\textwidth]{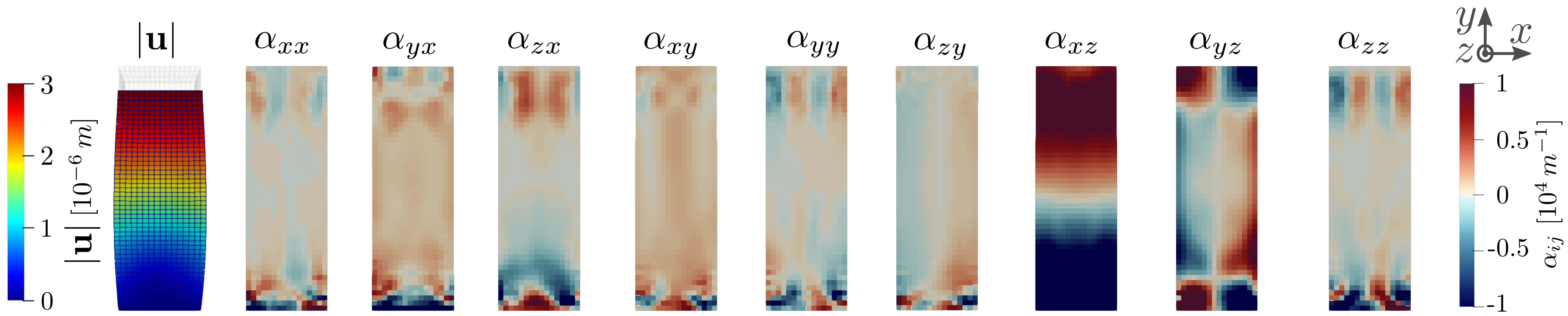}
    \caption{Deformed micropillar (size of $10\, \mu\mathrm{m}$) and all nine components of the Nye tensor at the micropillar surface for $\varepsilon_{yy} = -10\,\%$ mapped to the initial configuration. The last three alpha-components ($\alpha_{iz}$) are also determined in the experiment.
    }
    \label{fig:3c_alpha}
\end{figure*}

Caused by the compression loading, the micorpillar shows a characteristic deformation.
The resulting components of the Nye tensor are shown in Figure \ref{fig:3c_alpha}. 
Here $\alpha_{xz}$ and $\alpha_{yz}$ have the highest values and show a gradient along the micropillar height (y-direction) and micropillar width (x-direction), respectively.  
In addition to the three alpha components of the z-plane, that can be determined by the experiment, the simulation results also provide the remaining components of the complete alpha tensor.
Taking into account that it is not possible to avoid slight imperfections in experimental studies, we studied a number of possible imperfections that could have an impact on the results. 
The simulation results of the considered geometric imperfections of the micropillar are given in the Appendix A in Figure \ref{fig:7c_alpha_geo} showing the resulting alpha tensors.

\section{Discussion}
\label{sec:Discussion}

The objective of the present paper is to get further insight on the plastic material behavior and underlying dislocation microstrucuture evolution in the dislocation storage regime of fcc metals.
Therefore, single crystalline copper micropillars with a $\langle1\,1\,0\rangle$ crystal orientation and varying sizes between $1$ to $10\,\mu$m have been investigated under compression loading by experiments using \emph{in situ} HR-EBSD analysis as well as continuum dislocation dynamics simulations.

The plastic deformation of the micropillars occurs mainly on the four slip systems with the highest Schmid factors: $\{B2,B4,C1,C5\}$, as was to be expected. 
This can be observed in the experiment based on the deformation patterns and surface steps of the deformed micropillars (cp. the secondary electron image of the micropillar in Figure \ref{fig:sigmacomp}), taking into account the slip plane orientation (see Figure \ref{fig:2a_system}). 
This is consistent with the simulation results, that enable the determination of the allocation of plastic shear to the different slip systems (see Figure \ref{fig:3c_struc_slip}). 
Based on the superposition of the activity on all four primary slip systems (see Section \ref{ssec:2a_System}), the resulting $\vec e_x \otimes \vec e_x$ component of the plastic distortion tensor explains the much stronger bulging of the micropillar in x-direction than in z-direction in the simulation.

During loading, an increasing amount of GND density can be detected in both experiment and simulation for all micropillar sizes, see Figure \ref{fig:sigma-epsilon} and Figure \ref{fig:3c_struc_slip}.
This is consistent with the findings of Kiener \emph{et al.}, investigating the compression of single crystalline copper micropillars with a high symmetry crystal orientation via \emph{in situ} experiments and 2d discrete dislocation simulations \cite{kiener2011work}, and the findings of Maaß \emph{et al.}, using \emph{in situ} Laue and electron backscatter diffraction to examine the compression of single crystalline copper micropillars orientated for double slip \cite{maass2008crystal}.
The amount of GND density might be regarded as unintuitive for a homogeneous uniaxial loading of a single crystalline micropillar and raises the question on which slip systems the GND density is formed. And moreover, how the GND density is stabilized in the system.

The investigations presented in this paper show that the GND density is formed on the inactive slip systems for the compressed micropillars. The experimental results, see Figure \ref{fig:sigmacomp}, as well as the simulation results, see Figure \ref{fig:3c_struc_alloc}, show, that the allocation of the GND density is dominated by slip systems that are not mainly responsible for the production of plastic slip. This observation has also reported by Kirchlechner \emph{et al.} who considered tension tests of single crystalline copper specimen orientated for single slip and presented analyses of \emph{in situ} scanning electron microscopy and \emph{post mortem} $\mu$Laue diffraction \cite{kirchlechner2011dislocation}.
The simulation results in Figure \ref{fig:3c_struc_alloc} also show that there is an amount of GND density on active slip systems at the beginning of loading, but it is not preserved in the system during further loading and can leave the system via the surface.
In contrast, the fraction of the GND density on the inactive slip systems increases with increasing load and also as part of an increasing fraction of GND density of the total dislocation density within the system.
This can be explained by the fact, that the inactive slip systems have a significantly lower Schmid factor, which is zero for an ideally uniaxial external load. Therefore the driving force on the dislocations to leave the system is also significantly lower and the motion might be additionally hindered by existing network structures. Thus, the GND density on the inactive slip systems can easily be stabilized and preserved in the system. However, it should be remarked that the simulation does not resolve the GND and SSD parts in the network separately. Consequently, the network density could also contribute to the GND density in the system, which is not tracked according to the formulation used.

The Burgers vector analysis of the experimental data in Figure \ref{fig:acomp} identifies different regions and traces of dominant respective Burgers vectors. This yields a hint at the allocation of the experimentally obtained GND density to the individual slip systems. Obtained as an average Burgers vector this represents the  distribution of the GND density over the system as shown in Figure \ref{fig:alphacomp}.
Interestingly, the angle of traces (e.g. with Burgers vector 1) does not match the angle of the slip plane of the primary slip system (e.g. slip plane $C$ of slip system $C1$) but rather to the other active slip plane (e.g. slip plane $B$). 
From this we conclude that the GND density, or at least parts of it, is located on another inactive slip system. 
A superposition of active and inactive slip systems (e.g. $B4$ and $D6$) would be conceivable.
Glissile reactions between the moving dislocations on the active slip planes (e.g. $B4$) and the dislocations on the inactive slip systems (e.g. $D6$) can lead to new dislocations with the corresponding Burgers vectors (e.g. $D1$). However, the inactive slip plane (e.g. slip plane $D$) would run vertically in the cross section.

The resulting Nye tensor in the experiment and the simulation shows qualitatively good accordance in its individual components (cp. Figure \ref{fig:alphacomp} and Figure \ref{fig:3c_alpha}). 
The $\alpha_{xz}$ and $\alpha_{yz}$ components show gradients along the micropillar height ($y$-direction) and micropillar width ($x$-direction) respectively.  
Although there can be determined only three ($\alpha_{iz}$) of nine components in the experiment, the simulation shows that the most significant components are included.
However, it should be noted that in experiments at this length scale imperfections, such as, e.g., variations in the initial microstructure, misalignments, or unintended deformation states, can easily occur and impede or mislead the interpretation of the results. The imperfections can influence the measurements and the material behavior itself and thus lead to a certain scatter in the experimental results, as exemplary shown in Figure \ref{fig:allsizes} in Appendix \ref{imperfections} for the stress-strain curves for different pillar sizes. An essential advantage of the simulation is, that we can easily modify system and loading characteristics. So even if we don't know whether or not imperfections occur in the experimental results, a systematic investigation of possible imperfections by simulative variations can enable a better interpretation and strengthen confidence in the results. Comparing the experimental results in Figure \ref{fig:alphacomp} with the simulation data of the perfect system in Figure \ref{fig:3c_alpha}, we see, that there is qualitatively a good accordance between the experiment and the simulation. However slight imperfections leading e.g. to different gradients in the results can be better understood by taking into account the variations of the results due to imperfections shown in Figure \ref{fig:7c_alpha_geo}. A slight imperfection causing bending in the system might yield a change from the perfectly horizontal transition between positive and negative regions in $\alpha_{xz}$ (see Figure \ref{fig:3c_alpha}) to an inclined transition as shown for the bending case in Figure \ref{fig:7c_alpha_geo} and maybe slightly present in the experimental results.

Considering the dislocation microstructure in the system, it can be observed that the mobile dislocations partly leave the system via the surfaces or they interact with other dislocations in the micropillar and form a complex network structure. 
Which of these two processes is more pronounced is related to the size of the micropillar.
The transition from dislocation accumulation forming a cellular structure to the dislocation starvation regime can be observed both in the experiment (cp. Figure \ref{fig:alphacomp} and Figure \ref{fig:3um}) and in the simulation (see Figure \ref{fig:3c_size_sig}) as the pillar approaches smaller sizes.
In the considered size regime, the dislocation network formation is emphasized with increasing micropillar size for small strain states, as shown in Figure \ref{fig:acomp}, and decreases for smaller sizes as shown in Figure  \ref{fig:3c_size_sig}. 
This is not distinctively observable any more for larger strains in the simulation. 
However, the chosen time resolution in the simulation results might affect this observation since the time step could be chosen too large to capture all possible reactions during the ongoing density evolution. This remains to be investigated. 

For larger micropillars the probability of network formation increases due to the longer distance a dislocation has to travel to the surface as well as the lower dislocation velocities that can be observed.
The formation of the dislocation network structure contributes to the hardening, which can be observed in the stress-strain curves in both the experiments (Figure \ref{fig:sigma-epsilon}) and the simulations (Figure \ref{fig:3c_size_sig}).
This is consistent to the findings of Zhao \emph{et al.} analysing uniaxial compression of cylindrical copper micropillars with a $\langle1\,1\,0\rangle$ crystal orientation via scanning electron microscopy and scanning transmission electron microscopy \cite{zhao2019critical}.
However, considering the quantitative differences in the stress values between the experimental and the simulation results, it becomes clear that some mechanisms are not fully covered in the simulation model for the considered systems. So further adaptions of the used formulations, particularly regarding the reaction and interaction mechanisms with respect to the crystal orientation, have to be pursued.

\section{Conclusions}
\label{sec:Conclusions}

In this paper, we investigated the dislocation microstructure evolution in the dislocation storage regime of fcc metals.
Single crystalline copper micropillars with a $\langle1\,1\,0\rangle$ crystal orientation and varying sizes
between $1$ to $10\,\mu\mathrm{m}$ have been analysed under compression loading. The analysis comprised experimental in situ HR-EBSD measurements as well as 3d continuum dislocation dynamics simulations. The experimental setup provides insights into the material deformation and evolution of dislocation structures during continuous loading. This is complemented by the simulation of the dislocation density evolution considering dislocation dynamics, interactions, and reactions of the individual slip systems and providing direct access to these quantities.

It has been shown, that the plastic deformation of the material is mainly caused by dislocation activities on the four slip systems with the highest Schmid factor.
Here, the loss of dislocation density over the surface leads to observable deformation patterns and surface steps.
An interesting finding is the increasing amount of GND density in the system during loading. It has been observed for all micropillars considered and is located dominantly on the slip systems that are not mainly responsible for the production of plastic slip.
The stabilization of the GND density on these inactive slip systems can be explained by the lower Schmid factor and the associated lower driving force on the dislocations on these systems. However, the presented results indicate that the obstruction for dislocation motion caused by dislocation networks formation may play a significant role for this finding.

In the considered size regime, the dislocation network formation is emphasized with increasing micropillar size for small loading states. 
This indicates a transition from a dislocation starvation regime to a regime dominated by dislocation accumulation forming a cellular structure. 
This is accompanied by longer travel distance of dislocations to the surface as well as the lower dislocation velocities that can be observed for larger micropillars enforcing the probability of network formation and yielding a contribution to plastic hardening.

\section{Acknowledgements}
\label{sec:Acknowledgements}

The Authors appreciate the support during the setup of the in situ tests of Nicol\`{o} M. Della Ventura (Empa) and Xavier Maeder (Empa) and thank for discussions with Istv\'an Groma. SK was supported by the EPMAPOSTDOCS-II programme, that received funding from the European Union's Horizon 2020 research and innovation programme under the Marie Skłodowska-Curie grant agreement number 754364. This work was completed in the ELTE Institutional Excellence Program (1783-3/2018/FEKUTSRAT) supported by the Hungarian Ministry of Human Capacities. PDI acknowledges the support of the National Research, Development and Innovation Fund of Hungary (contract number: NKFIH-K-119561). The simulation work was performed on the computational resource ForHLR II funded by the Ministry of Science, Research and the Arts Baden-Württemberg and German Research Foundation (DFG). KZ and KS acknowledge the financial support for the research group 1650 by DFG under contract number GU367/36-2. 

\appendix
\label{sec:Appendix}

\section{Impact of geometric imperfections}
\label{imperfections}

\begin{figure*}[!ht]
    \centering
    \includegraphics[width=0.45\textwidth]{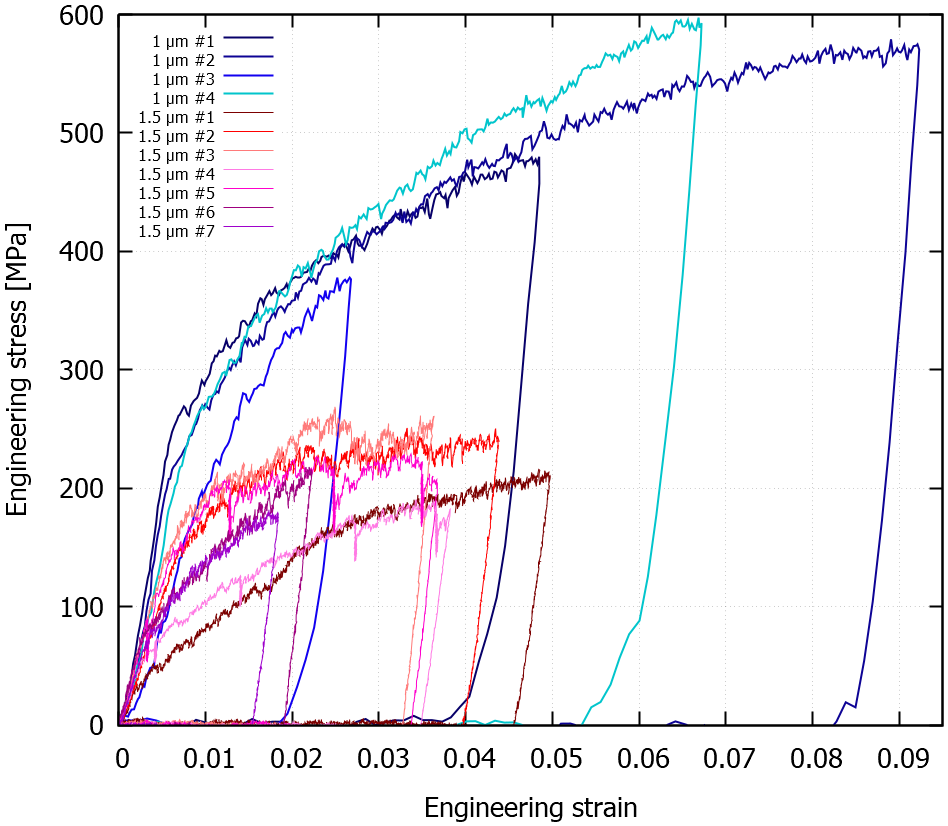}
    \hspace{0.1cm}
    \includegraphics[width=0.45\textwidth]{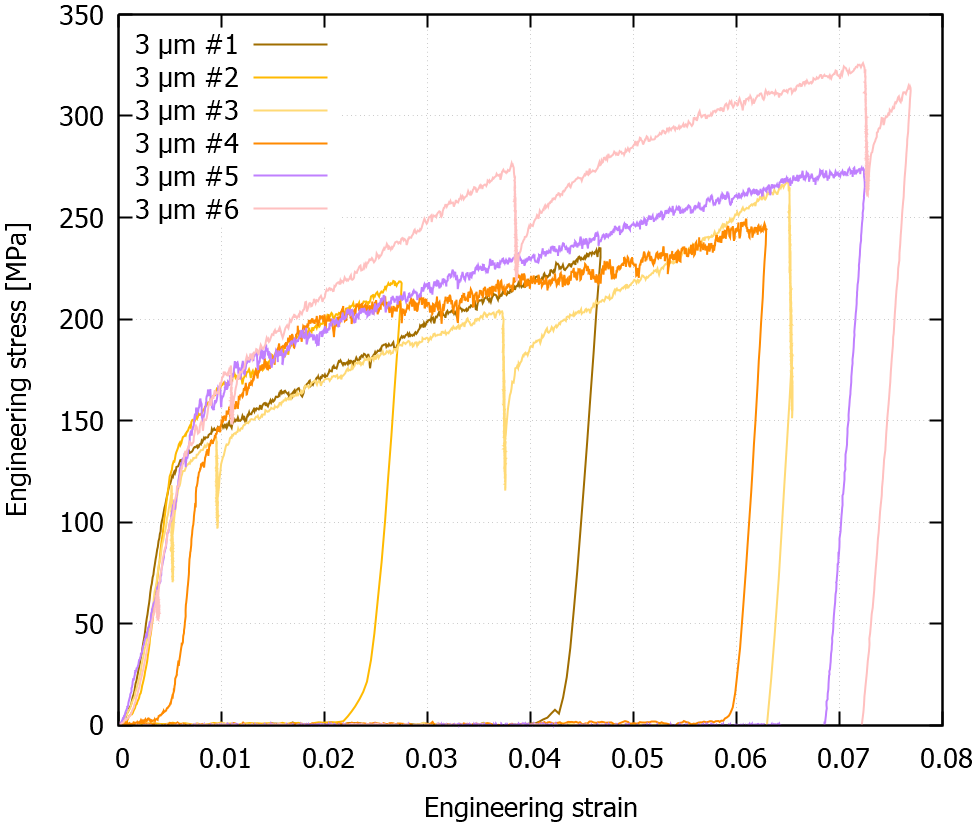}
    \hspace{0.1cm}
    \includegraphics[width=0.45\textwidth]{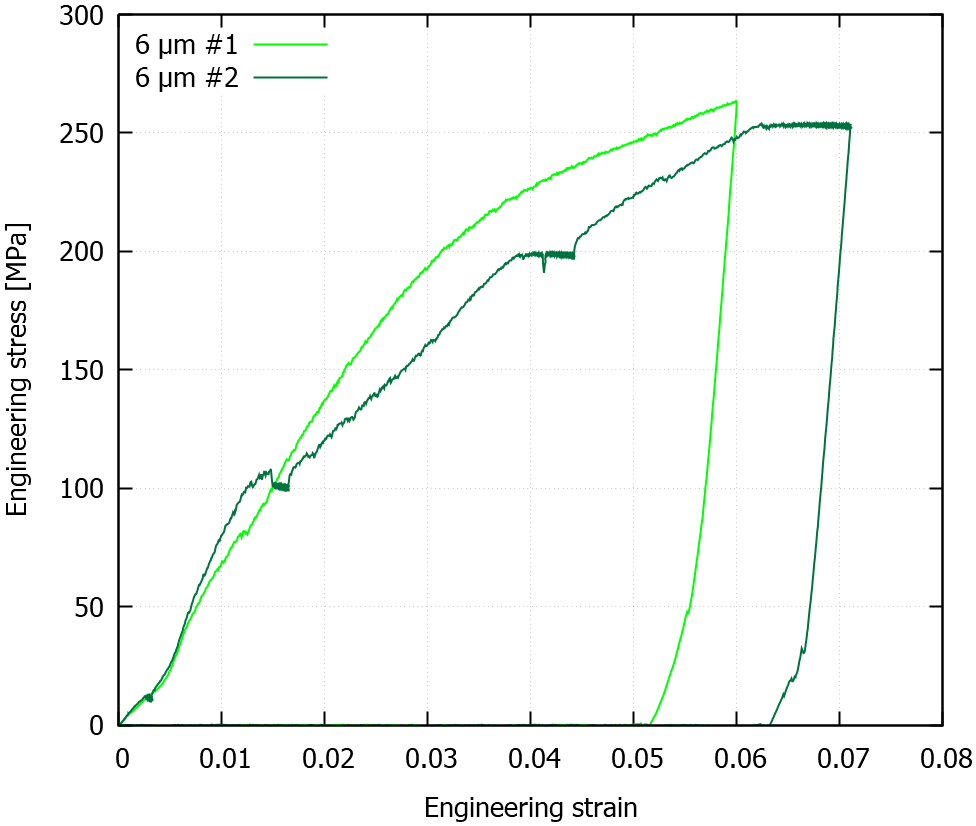}
    \hspace{0.1cm}
    \includegraphics[width=0.45\textwidth]{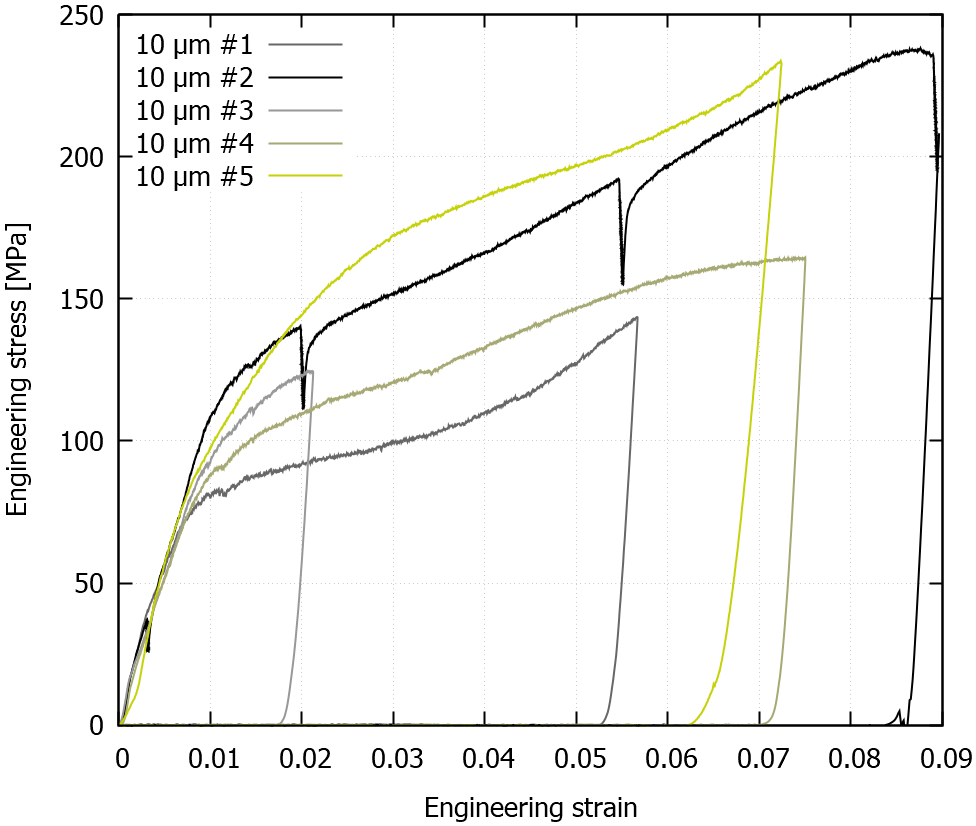}
    \caption{Engineering stress-strain curves for different sized square pillars (1, 1.5, 3, 6, and 10 $\mu$m side length). 6 $\mu$m pillars were bent due to misalignment of the flat punch. }
    \label{fig:allsizes}
\end{figure*}

Numerous micropillars were compressed experimentally in this study. However, it is well known that to achieve a perfect system and  deformation process without imperfections is a quite difficult task. Any deviation from perfect alignment or minor sample preparation differences might have an impact on the results, e.g., the recorded $\sigma(\epsilon)$ curves will differ from each other. Although pillars were prepared sequentially ensuring good similarity, they can't be a perfect replica of one another and, e.g., the initial dislocation density can vary from place to place. The indenter flat punch tip can have a slight misalignment with the pillar's top surface, which will enhance bending upon deformation. For completeness of the investigations shown in this paper, we show all corrected  $\sigma(\epsilon)$ curves of the experimental analyses plotted in Figure \ref{fig:allsizes}.

In case of the 6 $\mu$m pillars the flat punch tip was not well aligned with the top surface of the specimen, hence the change in the steepness of $\sigma(\epsilon)$ at $\epsilon \approx 0.005$. At the early onset of the deformation process, in case of sample--tip misalignment, the crystal is rotated in order to accommodate the mismatch. Afterwards, the sample's $\sigma(\epsilon)$ curve is modified, as more GNDs are introduced into the system through this geometric imperfection, leading to a less pronounced elastic-plastic transition in the stress-strain curve. It needs to be noted that one of the 6 $\mu$m pillars was deformed unlike the other in situ tests. When pausing the deformation of this pillar, instead of keeping the displacement constant (like i.e. the 10 $\mu$m pillar case) the load was kept constant, hence the difference in the character of the $\sigma(\epsilon)$ curve's shape. In order to avoid any excess slip caused by this continuous punch tip movement, the subsequent in situ experiments were suspended by the "constant displacement" method.

\begin{figure*}[!ht]
    \centering
    \includegraphics[width=0.95\textwidth]{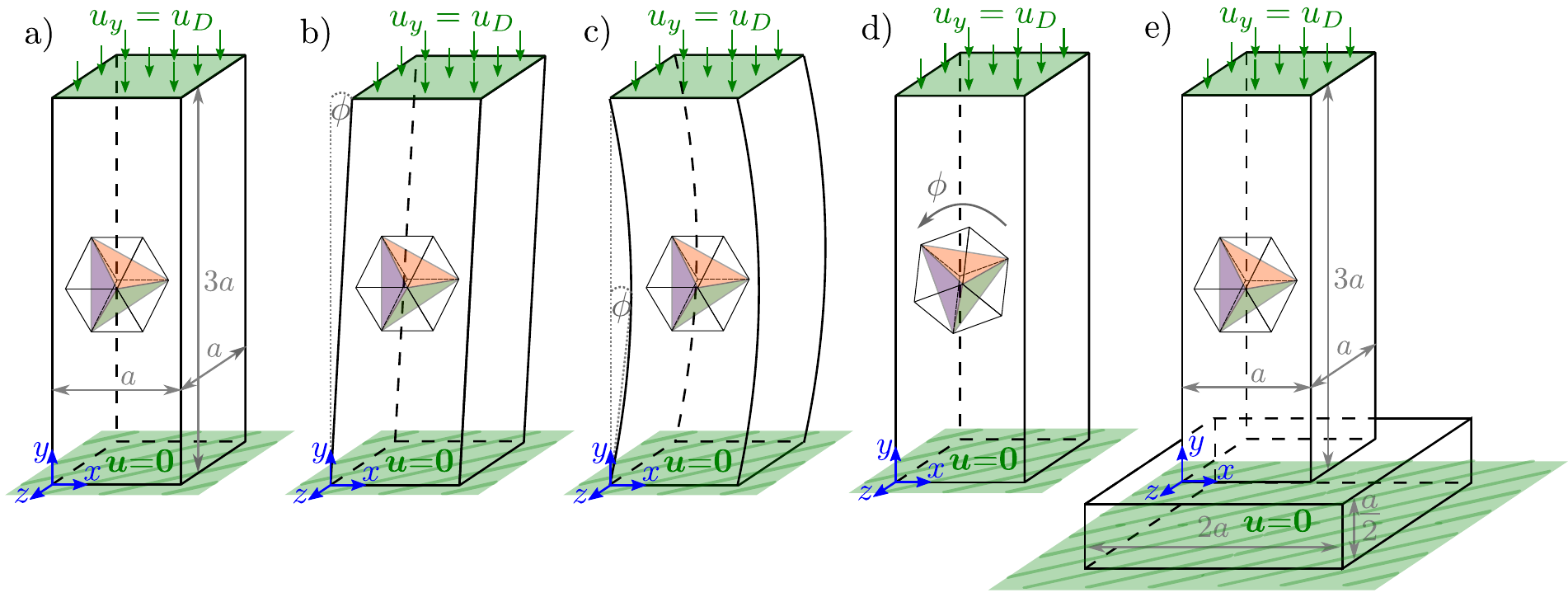}
    \caption{Modelling of geometric imperfections for a micropillar of $a=6\,\mu m$ size. The angle is chosen to $\theta=4^\circ$.}
    \label{fig:7c_system_geo}
\end{figure*}

An essential advantage of the simulation is, that we can easily modify system and loading characteristics. So even if we don't know whether or not there occur imperfections in the experimental results, a systematic investigation of possible imperfection by simulative variations can enable a better interpretation of the results. The effects of various imperfections on the spatial distribution of the Nye tensor are studied by simulation in this section.
Starting from an ideally straight micropillar as a reference system (system a), the following geometric imperfections are considered, as shown in Figure \ref{fig:7c_system_geo}. A linear shift of the micropillar in z-direction (system b), a curved bow-out of the micropillar in z-direction (system c), and a misalignment in the crystal orientation around the z-axis (system d).
The linear shifted and curved micropillars (system b and c) show the same offset in z-direction at half height.
Additionally, an embedded micropillar (system e) is considered to investigate the influence of simplifying assumptions in the simulation compared to the experiment. It should be noted that the sharp notch in the material in the embedded system is a numerical challenge and can lead to inaccuracies.

\begin{figure*}[!ht]
    \centering
    \includegraphics[width=0.995\textwidth]{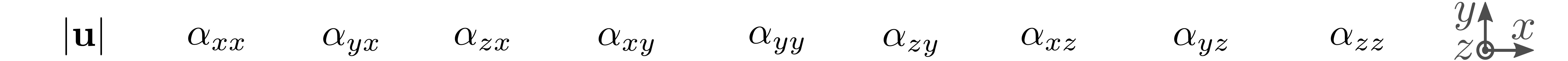}
    \includegraphics[width=0.9\textwidth]{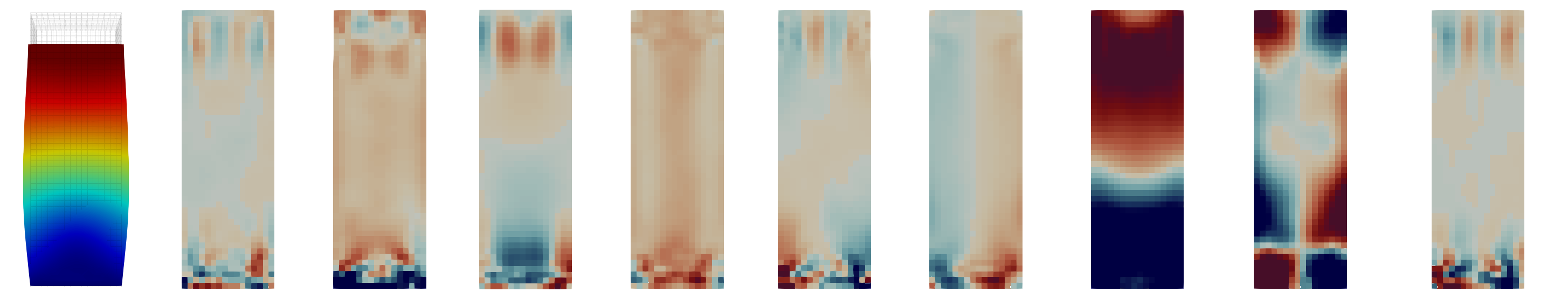}
    \includegraphics[width=0.065\textwidth]{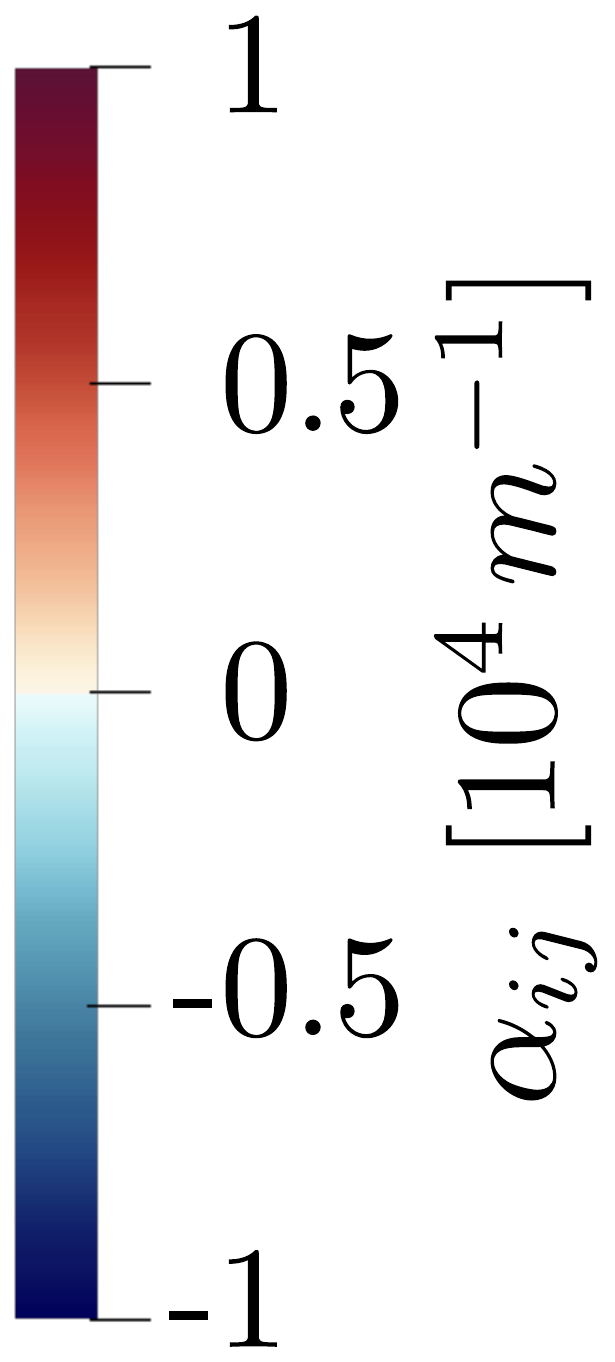}
    \includegraphics[width=0.9\textwidth]{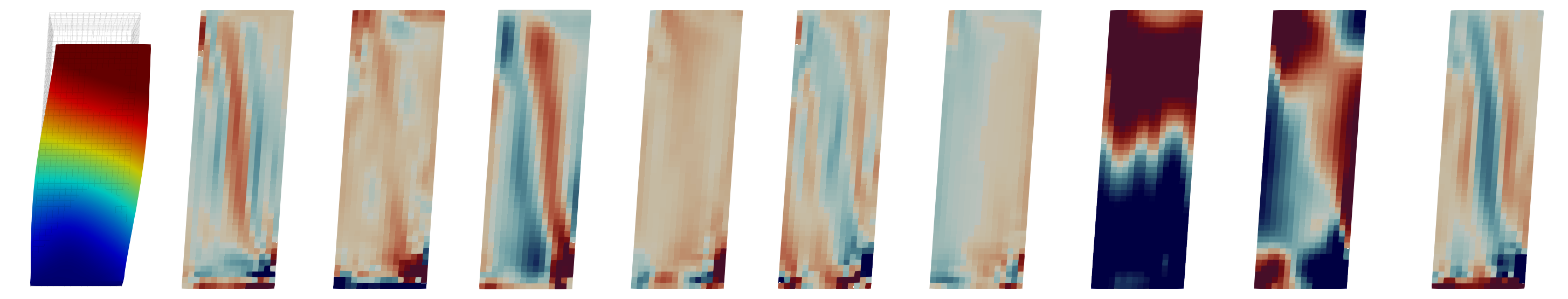}
    \includegraphics[width=0.065\textwidth]{scale_alpha_el.pdf}
    \includegraphics[width=0.9\textwidth]{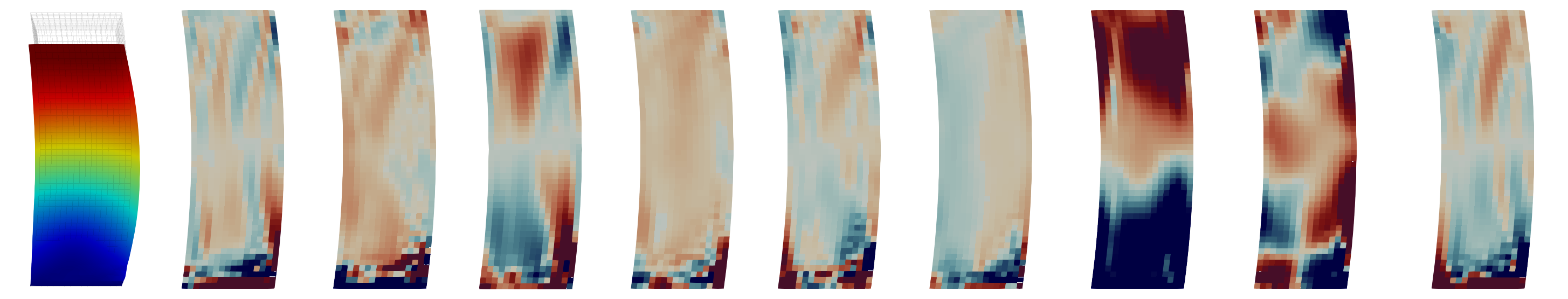}
    \includegraphics[width=0.065\textwidth]{scale_alpha_el.pdf}
    \includegraphics[width=0.9\textwidth]{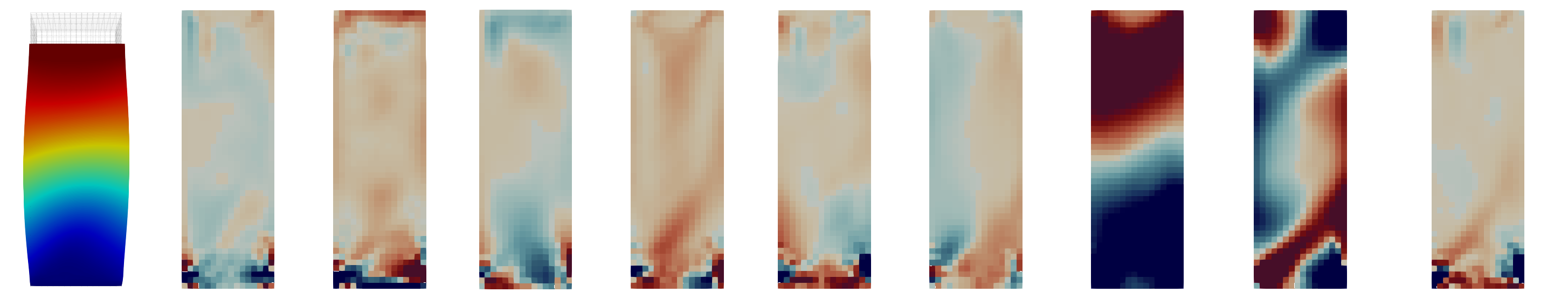}
    \includegraphics[width=0.065\textwidth]{scale_alpha_el.pdf}
    \includegraphics[width=0.9\textwidth]{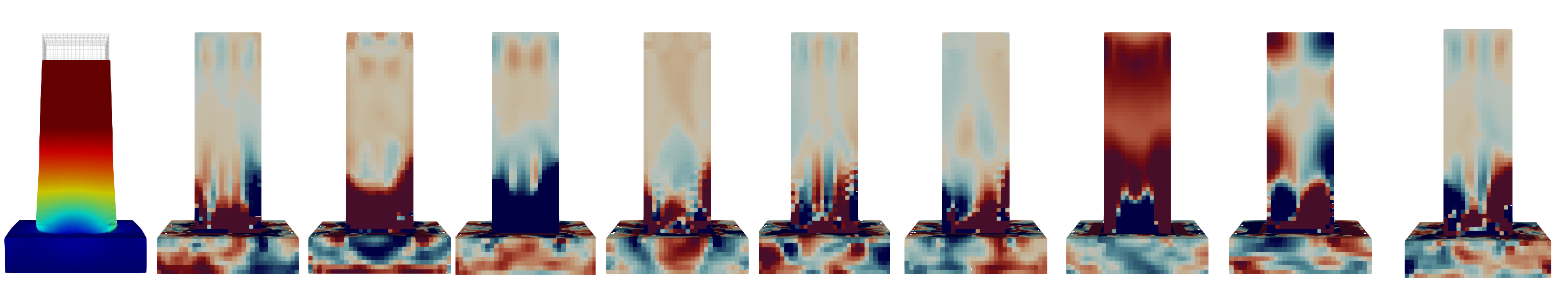}
    \includegraphics[width=0.065\textwidth]{scale_alpha_el.pdf}
    \caption{Absolute displacements of the deformed micropillar (scaling factor of two) and all nine components of the Nye tensor mapped to the initial configuration at a loading of $\varepsilon_{yy} = -5.7\,\%$.  
    } 
    \label{fig:7c_alpha_geo}
\end{figure*}

The effects of the geometric imperfections on the deformed state and the Nye tensor field are shown in Figure \ref{fig:7c_alpha_geo}.
The qualitative distribution of $\alpha_{zx}$ and $\alpha_{zy}$ components is similar in all systems, but the angles and size of the transition regions change. For the linear shifted micropillar (system b), a characteristic structure is additionally observed in the $\alpha_{zz}$ component.
%


\bibliographystyle{crplain}

\nocite{*}

\bibliography{8_literature}

\end{document}